\documentclass{aastex61}

\usepackage{threeparttable}
\usepackage{multirow}
\usepackage{lineno}

\accepted{May 24, 2018}
\submitjournal{PASP}

\shorttitle{New discovered DBWDs in SDSS DR12 and DR14 }
\shortauthors{Kong X., Luo A.-L., Li X.-R., Wang Y.-F., Li Y. \& Zhao J.-K.}

\begin{document}

\title{Spectral feature extraction for DB White Dwarfs through Machine Learning applied to new discoveries in the Sdss DR12 and DR14}

\correspondingauthor{A-Li Luo}
\email{lal@nao.cas.cn}

\author[0000-0001-8011-8401]{Xiao Kong}
\affil{Key Laboratory of Optical Astronomy, National Astronomical Observatories, Chinese Academy of Sciences, \\
Beijing 100012, China}
\affiliation{University of Chinese Academy of Sciences, \\
Beijing 100049, China}

\author[0000-0001-7865-2648]{A-Li Luo}
\affiliation{Key Laboratory of Optical Astronomy, National Astronomical Observatories, Chinese Academy of Sciences, \\
Beijing 100012, China}
\affiliation{University of Chinese Academy of Sciences, \\
Beijing 100049, China}

\author{Xiang-Ru Li}
\affiliation{School of mathematical Sciences, South China Normal University, \\
Guangzhou, 510631, China}

\author{You-Fen Wang}
\affiliation{Key Laboratory of Optical Astronomy, National Astronomical Observatories, Chinese Academy of Sciences, \\
Beijing 100012, China}

\author{Yin-Bi Li}
\affiliation{Key Laboratory of Optical Astronomy, National Astronomical Observatories, Chinese Academy of Sciences, \\
Beijing 100012, China}

\author{Jing-Kun Zhao}
\affiliation{Key Laboratory of Optical Astronomy, National Astronomical Observatories, Chinese Academy of Sciences, \\
Beijing 100012, China}



\begin{abstract}

Using a machine learning (ML) method, we mine DB white dwarfs (DBWDs) from the Sloan Digital Sky Survey (SDSS) Data Release (DR) 12 and DR14.
The ML method consists of two parts: feature extraction and classification.
The least absolute shrinkage and selection operator (LASSO) is used for the spectral feature extraction by comparing high quality data of a positive sample group with negative sample groups.
In both the training and testing sets, the positive sample group is composed of a selection of 300 known DBWDs, while the negative sample groups are obtained from all types of SDSS spectra.
In the space of the LASSO detected features, a support vector machine is then employed to build classifiers that are used to separate the DBWDs from the non-DBWDs for each individual type.
Depending on the classifiers, the DBWD candidates are selected from the entire SDSS dataset.
After visual inspection, 2808 spectra (2029 objects) are spectroscopically confirmed. By checking the samples with the literature, there are 58 objects with 60 spectra that are newly identified, including a newly discovered AM CVn.
Finally, we measure their effective temperatures ($T_{\text{eff}}$), surface gravities (log $g$), and radial velocities, before compiling them into a catalog.

\end{abstract}




\section{Introduction}

At the final stage of stellar evolution for main sequence stars, white dwarfs (WDs) simply cool off in the absence of nuclear reactions. The energy of most WDs are generated by the radiation of the residual gravitational contraction, instead of nuclear fusion. Generally, the initial masses of the progenitors of WDs are approximately between 0.07 and 8 $M_{\odot}$,  and their radius are often the same order as that of the Earth, implying that they need extremely long cooling times. It is believed that over 97\% of the stars in the Galaxy will eventually end up as WDs \citep{2001PASP..113..409F}.
The luminosity function of a WD, containing information of the stellar death rate in the local galactic disk, can be used to estimate the density of the matter in the Galaxy. A statistically complete sample is required to measure the luminosity function of WDs \citep{2010ApJ...714.1037L}.

Approximately 80\% of all observed WDs belong to DA type with Hydrogen dominated atmospheres, with the remaining 20\% falling into the DB (He {\small \bf I}) or DO (He {\small \bf II}) categories with atmospheres dominated by helium.
As these stars are lined up in the WD cooling sequence, they are observed with temperatures of approximately 45,000 K; categorized as hot DO stars with He {\small \bf II} rich in spectra, and of effective temperatures ($T_{\text{eff}}$) mostly below 30,000 K; categorized as DB WDs (DBWDs) with only He {\small \bf I} lines in the spectra. When the temperature drops to 10,000 K, helium becomes spectroscopically invisible, e.g., featureless smooth DC, carbon present DQ, or metal rich DZ spectra \citep{2007A&A...470.1079V}.

With nearly pure helium in the neutral form in their atmospheres, the DBWDs represent the best example of hydrogen-deficient stars in the universe. 
Many hydrogen dominated DA WDs transform into DBWDs with helium atmospheres, and the ratio of DA to non-DA WDs varies as a function of $T_{\text{eff}}$ along the cooling sequence \citep{2001PASP..113..409F}. By expanding the DBWD sample size, a better understanding of the evolution of WDs is possible.

The high photospheric purity of DBWDs was firstly revealed by atmosphere models \citep{1970A&A.....7...91B}. Only about 80 optical spectra and 25 ultraviolet spectrophotometries were investigated in the 1900s \citep{1996ASPC...96..295B}. 
Later, with the help of the Sloan Digital Sky Survey (SDSS) spectral surveys, systematic searches were completed and a larger sample of DBWDs were obtained, holding great potential for the exploration of the chemical evolution of DB degenerates.
\cite{2013ApJS..204....5K} provided 922 DBWDs from SDSS Data Release (DR) 7 \citep{2009ApJS..182..543A}, \cite{2015MNRAS.446.4078K} added another 450 in DR10 \citep{2011AJ....142...72E}.
DR12 \citep{2015ApJS..219...12A} increased this number by 121 \citep{2016MNRAS.455.3413K}, among which \cite{2015A&A...583A..86K} selected 1267 spectra with signal-to-noise ratios (S/N) greater than 10, and of these the atmospheric parameters of 1107 objects were analyzed.
Based on the sample of 150 DBWDs and 1733 DA WDs, \cite{2007MNRAS.375.1315K} provided the average masses of $0.711 \pm 0.009 M_{\odot}$ and $0.593 \pm 0.016 M_{\odot}$ for the DB and DA types, respectively.
\cite{2015A&A...583A..86K} reported that the mass of DB types have a significant increase below $T_{\text{eff}}=$ 16,000 K, possibly caused by the imperfect implementation of line broadening of neutral helium atoms, and analyzed the distributions of DBA WDs and DBWDs with the height, $z$, above the Galactic plane differing toward lower $T_{\text{eff}}$.
\cite{2006AJ....132..676E} presented 28 stars as candidate hot DB or cool DO WDs, some of which are the first helium atmosphere WDs found in the range 30,000--45,000 K, in the DB gap.

However, the majority of known DB spectra are obtained through parameter measurement, which may lead to incompleteness in the DBWD findings because of bad data quality or spectral fitting failures.  For such a large data set from SDSS DR12,  including 4355,200 spectra, some DBWDs will not be identified, especially for data with low S/N. The DB class in the SDSS catalog is lacking as most known DBWDs are mis-classified as ``O,'' ``B,'' ``A,'' ``QSO'' or some other types in the DR12.  
It should be noted that there is a ``WD'' class in SDSS DR12, which mainly includes DA WDs, although a few other subtypes of WD are mixed in.

We search for DBWDs in all DR12 and DR14 without color or other limits, which depend on machine learning (ML). 
After a first manual check, we discard those without obvious He {\small \bf I} lines (4471.5 and 5875.6\AA). Then we arrange the remaining spectra in descending order of S/N of the g band (S/N\_g), where the majority of He {\small \bf I} lines exist, and select the top 300 as our positive samples, which is used to extract the DB features using the LASSO method \citep{Tibshirani96regressionshrinkage}.
To analyze the features, the mathematical tool Wavelet \citep{1992wavelet} is used to check the decomposed spectra in different scales.
The wavelet transform is able to cut up signals into different frequency components, then each component with a resolution matched to its scale can be studied. 
We directly employ the two built-in functions of MATLAB, ``wavedec'' and ``wrcoef,'' to aid the wavelet analysis of the spectrum.
\cite{2015ApJS..218....3L} described the basic properties of wavelet, explained the process of wavelet decomposition, and  employed it in combination with LASSO to estimate stellar atmospheric parameters. \cite{2001NCimB.116..879L} used it to obtain spectral classification information of galaxies.
Afterwards, in the derived feature space rather than the original spectra, SVM is employed to distinguish DB from other types of objects.
``Feature'' here has the same meaning as flux at some particular wavelength or any specific location of spectra.

This paper is organized as follows. 
Section \ref{sec:data} describes the spectral data used in this paper. 
The ML method applied in this paper are explained in detail in Section \ref{sec:method}, including the preprocessing of the data set, feature model construction of DBWD through the LASSO algorithm, feature analysis using the wavelet transform, and classifier establishment by the SVM. 
Then, we apply the method to detect DBWDs from the SDSS dataset, as introduced in Section \ref{sec:recognition}. 
In Section \ref{sec:analysis}, we compare our results with those of the literature and calculate the relative parameters, such as $T_{\rm eff}$, surface gravities (log $g$), and ultra-violet color ($\text{FUV} - \text{NUV}$).
Finally, we summarize our work in Section \ref{sec:conclusion}.

\section{Data Sets}
\label{sec:data}
In the \cite{2015A&A...583A..86K} catalog, there are a total of 1107 objects of DBWD from SDSS DR10 and DR12, which are originally classified as O, QSO, B, WD, galaxy, or other types by SDSS 1D pipeline; the quantities of each type are given in Table \ref{tab:ori}. 
This table suggests that we can pick out more DB spectra from all types of SDSS spectra, excluding those with a high confidence of classification.

\startlongtable
\begin{deluxetable}{ccr|ccr|ccr}
\tablecaption{Classfication and quantities of 1107 DBWDs in the SDSS SR12 catalog. \label{tab:ori}}
\tablehead{
\colhead{Class\tablenotemark{a}} & \colhead{Subclass\tablenotemark{a}} & \colhead{Number} &
\colhead{Class\tablenotemark{a}} & \colhead{Subclass\tablenotemark{a}} & \colhead{Number} &
\colhead{Class\tablenotemark{a}} & \colhead{Subclass\tablenotemark{a}} & \colhead{Number}
}
\startdata
QSO		&	null		&	416		&	galaxy	&	null	&	12	&	star	&	K			&	2	\\
Star	&	OB			&	400		&	star	&	A		&	11	&	star	&	T			&	2	\\
Star	&	O			&	117		&	star	&	CV		&	9	&	star	&	carbon		&	2	\\
Star	&	B			&	62		&	star	&	F		&	7	&	star	&	G			&	1	\\
QSO		&	broadline	&	30		&	star	&	L		&	4	&	galaxy	&	broadline	&	1	\\
Star	&	WD			&	28		&	star	&	M		&	3	&	\nodata	&	\nodata		&	\nodata	\\
\enddata
\tablenotetext{a}{``Class'' and ``subclass'' are adopted from the data archive of SDSS.}
\end{deluxetable}

\begin{table}
\caption{Roles of the Three Data Sets.}
\label{tab:dataset}
\centering
\scriptsize
\begin{tabular}{ll} \hline
\multicolumn{1}{c}{Data Set} & \multicolumn{1}{c}{Roles} \\ \hline
Training Data & To be used in the training process, i.e.	\\
	&	Detecting features by LASSO (Section \ref{sec:feature});	\\
	&	Estimating the parameterizing model by LIBSVM (Section \ref{sec:trainsvm}).	\\
Testing Data & To be used in the training process, i.e.	\\
	&	Determining the parameters in LASSO (Section \ref{sec:feature});	\\
	&	Determining the hyper-planes in LIBSVM (Section \ref{sec:trainsvm}).	\\
Experimental Data	&	Application of Section \ref{sec:method}, to be used in \\
	&	Searching DB spectra from Experimental Data (Section \ref{sec:recognition}).	\\ \hline
\end{tabular}
\end{table}

We attempt a ML method to search for more DBWDs, with a focus on the low quality data. The basic idea is to use SVM as a classifier to sort out the DBWDs from the spectral data of SDSS DR12 and DR14, based on features found by LASSO.

We construct two subsets for training and testing from SDSS DR12. The training set is used for learning, that is to fit the parameters (features and hyper-planes) of the classifier. The testing set is used for the parameter adjustment of the classifier, e.g., to choose the best features and most suitable kernel function of SVM. A validation process of 10-fold cross-validation is built into both the LASSO and SVM packages and conducted automatically using the training set. We select a candidate data set, named experimental data (ED), from the SDSS DR14 and explain the procedure of selection in Section \ref{sec:datareco}. Table \ref{tab:dataset} lists the roles of the training, testing, and ED sets. 

\begin{table}
\caption{Types of spectra applied in the experiment.}
\label{tab:type}
\centering
\begin{threeparttable}
\begin{tabular}{cl} \hline
Class\tnote{a}	&	\multicolumn{1}{c}{Subclass\tnote{a}}	\\ \hline
Star	&	O, B, A, F, G, K, M, L, T, WD, carbon, CV		\\
Galaxy	&	AGN, broadline, starburst, starforming, null	\\
QSO		&	AGN, broadline, starburst, starforming, null	\\ \hline
\end{tabular}
\begin{tablenotes}
\item[a] ``Class'' and ``Subclass'' are adopted from the data archive of SDSS.
\end{tablenotes}
\end{threeparttable}
\end{table}

In the SDSS data archive, spectra were grouped into ``Class'' and ``Subclass,'' which are shown in Table \ref{tab:type}. As the basic SVM is a binary-classification algorithm, the classification between DB and each ``Subclass'' are performed in parallel. For convenience, we abbreviate ``Class + Subclass'' (CPS) as the ID of each subclass in this experiment, such as ``star+O'' or ``QSO+AGN.'' 

\subsection{Data for Training Process}
\label{sec:datatrain}

After a visual inspection, we select 300 DB samples of DBWDs with the highest S/N\_g as the positive samples of the training set. This is the only group of positive samples, which means it is compared with all groups of negative samples. Clearly, the redshift of the positive samples from the SDSS DR12 is incorrect because they are measured using non-DB templates. Hence, we need to re-measure their $z$ values using DB templates, and then move them to the rest frame.

Next, we apply full spectral template matching to accomplish this process.
This is the most-widely used method in spectral classification and measurements and is also the core algorithm in the ``one-dimensional'' pipeline software of SDSS \citep{2008AJ....136.2022L}.
It is approached as a $\chi^{2}$ minimization problem. 
We firstly reshape the pseudo continuums of the templates to ensure they are consistent with the spectrum, then calculate the distance between a template and the spectrum at each step within a specific redshift range.
Finally, $z$ can be derived from the template that has the minimum $\chi^{2}$, which is called the best-fit.

Meanwhile, for each CPS, the spectra selected from the total set of spectra, with the ranking of S/N\_g in a descending order, are the negative samples.
In the algorithm application, SVM has many limitations when it is applied to the binary-classification procedure from imbalanced datasets, in which the negative instances heavily outnumber the positive ones.
Although many applications have been raised to overcome this issue \citep{10.1007/978-3-540-30115-8_7}, we decided to keep a balance between positive and negative samples to ensure the correction of the classifications, meaning that only 300 spectral data are kept in each group of the negative samples.
Furthermore, five groups of each CPS are built as negative samples in order to obtain more comprehensive results; a total of 1500 spectra in every CPS in the Table \ref{tab:type}.

\subsection{Data for Recognition}
\label{sec:datareco}

DR12 and DR14 contain huge amounts of spectral data, most of which have a high quality and correct classification.
It would be inefficient if all spectra were included in the searching process, especially for ``GALAXY'' that corresponds to the largest number in the catalog.

\begin{figure}
\centering
\includegraphics[width=0.48\textwidth]{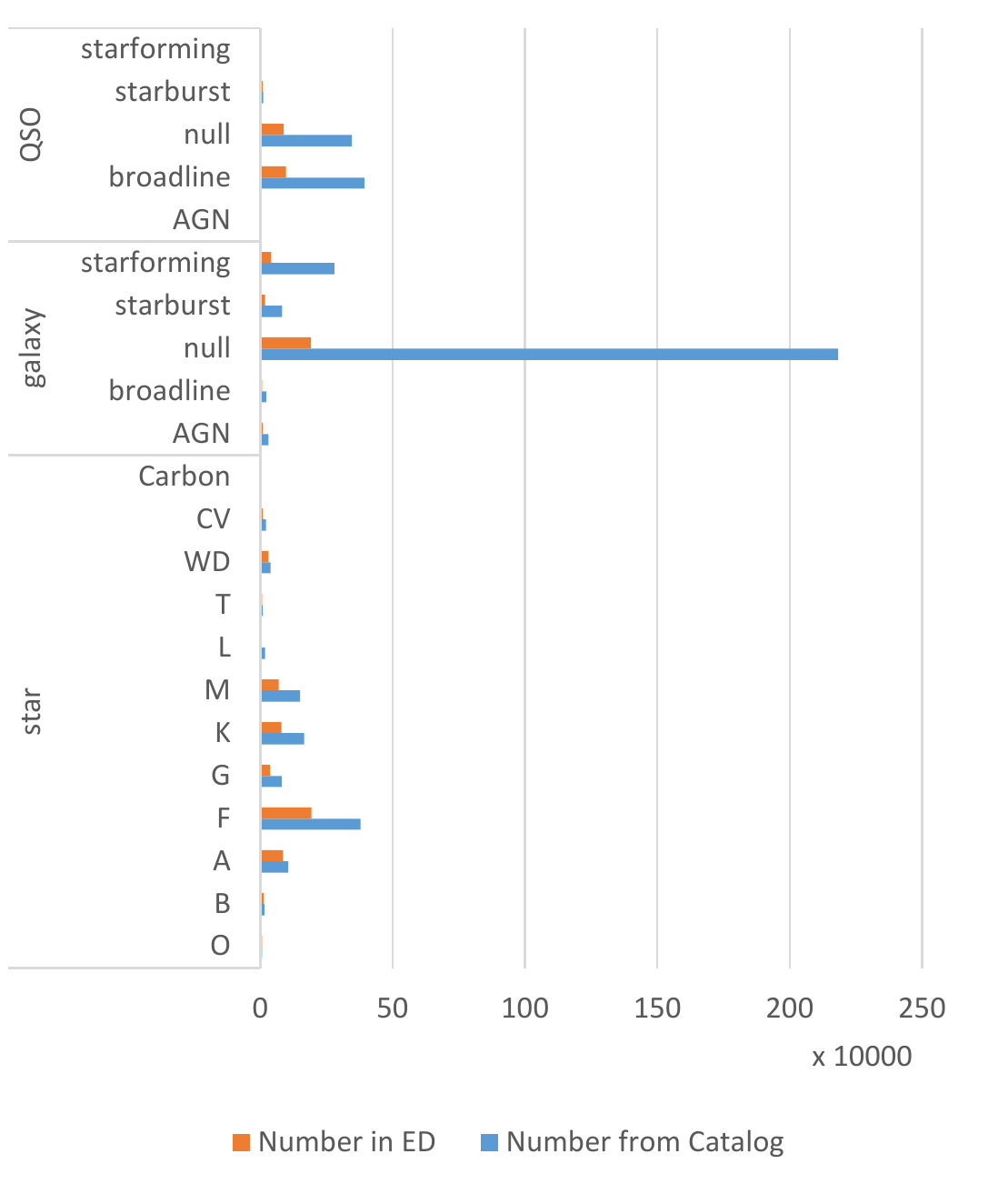}
\caption{Data usage of the experiment. The numbers of spectra from the SDSS DR14 catalog in each CPS are shown as the blue lines, while those from the ED are in red.}
\label{fig:ed}
\end{figure}

We adopt the full spectral template matching program to classify all spectra from the SDSS dataset.
DB templates are replaced by all templates of the LAMOST 1D Pipeline \citep{2015RAA....15.1095L}, which differs from the steps introduced in Section \ref{sec:datatrain}.
In order to obtain reliable classification results, the relationship of $\chi^{2}$ between the best and second-best fit is also taken into consideration.
After this preprocessing, spectral data that are not DBWDs with a high degree of confidence will be excluded, while others remain as ED, from which DBWD is recognized in Section \ref{sec:recognition}.
The amount of spectral data from each CPS within the ED and catalog are illustrated in Figure \ref{fig:ed}, using the red and blues bars, respectively.
Compared with the SDSS DR14 data set, the ED is eventually built by reducing the quantity by an average of 75 percent, after the reduction process.

\section{Method in the Training Process}
\label{sec:method}
\begin{figure}
\centering
\includegraphics[width=0.48\textwidth]{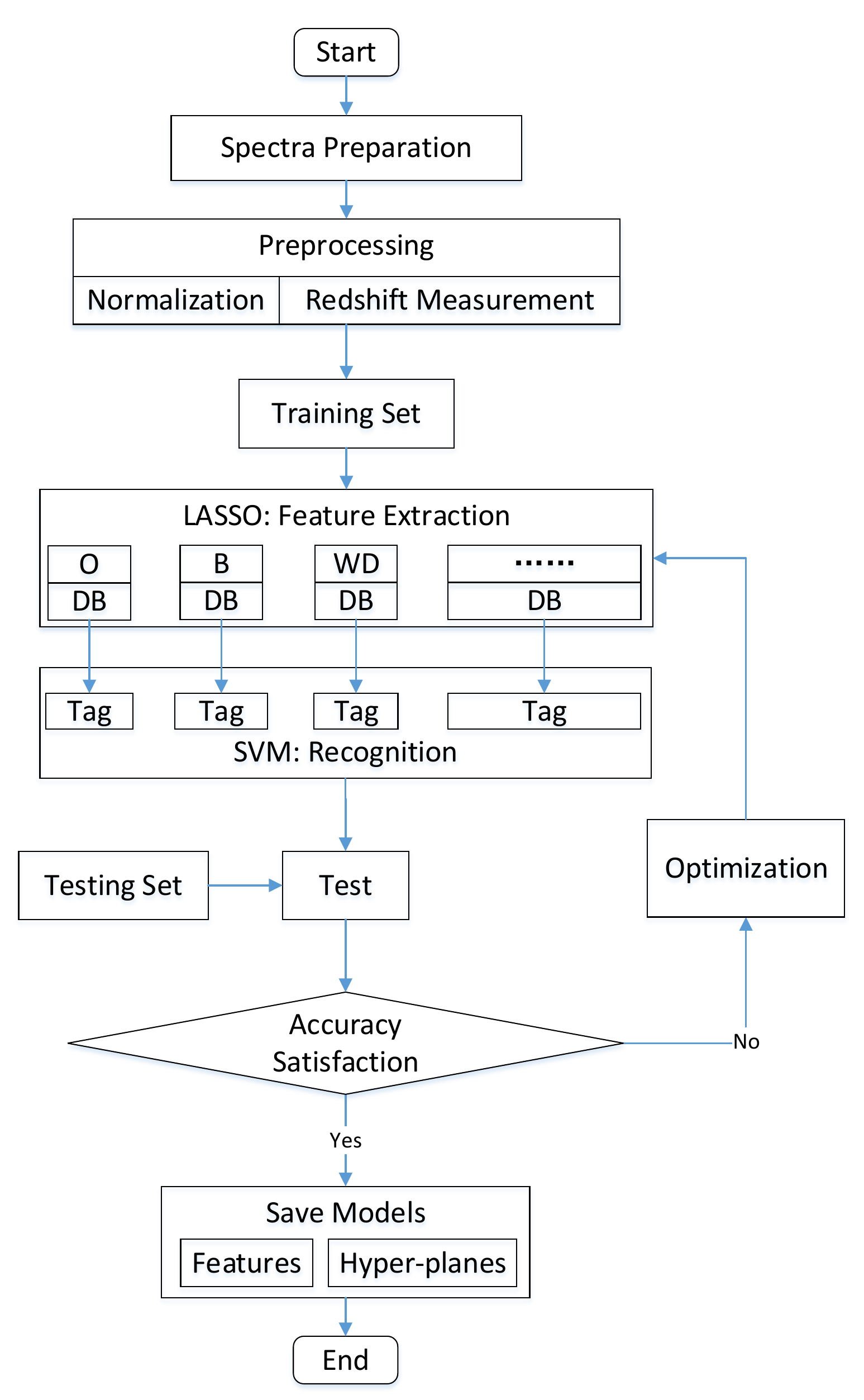}
\caption{Flowchart of the training process of this experiment, 
mainly consisting of preprocessing, feature extraction, classification, and optimization.}
\label{flch:all}
\end{figure}

The flowchart of the training process is given in Figure \ref{flch:all}. All of the prepared data begin with the preprocessing procedure, including the normalization and redshift measurement. The spectral features are then extracted through LASSO for each group, followed by a bi-classify modular of SVM to separate the DBWDs from all of the types. If the accuracy (see Section \ref{sec:verification} in detail) is not high enough, an optimization process needs to be performed, i.e., remove some contamination of DB from the negative sample sets and restart this loop. Once the training process is completed, the unique features of the DBWDs and hyper-planes of each group can be derived by LASSO and SVM, respectively. In addition, we analyze the features extracted by LASSO using Wavelet and provide a multi-scale explanation. 

\subsection{Data preprocessing}
\label{sec:preprocess}

\subsubsection{Normalization for Positive and Negative Samples}
To ensure the consistency of different spectral data, a simple normalization is required before feature extraction.
Let a vector $x = (x_{1}, x_{2}, \ldots, x_{n})^{T}$ represent a spectrum, where n ($n>0$) is the number of points.
The component $x_{i}$ represents the flux of the spectrum $x$, $i \in \{1,2,\ldots,n\}$. 
We simply put all of the flux between -1 and 1,
\begin{eqnarray*}
\hat{x} = \frac{x - \bar{x}}{\sigma_{x}}
\end{eqnarray*}
where $\bar{x}$ and $\sigma_{x}$ are the mean value and variance of $x$, respectively. 

\subsubsection{Redshift Measurement for Positive Sample Groups}
In our investigation, we find that most DB spectra from the SDSS dataset have incorrect redshifts, especially for those with large redshifts that are classified as  ``GALAXY'' and ``QSO'' in the catalog.
To ensure consistency of the feature extraction, all samples in the training set should be in the rest frame.

Firstly, the DBWD spectra with high quality are selected as DB templates that only used for the redshift measurements. Then, we apply full spectral template matching, which is described in the Section \ref{sec:datatrain}, to measure all samples in the training set.

\subsection{Feature Extraction}
\label{sec:feature}
\begin{figure}
\centering
\includegraphics[width=0.48\textwidth]{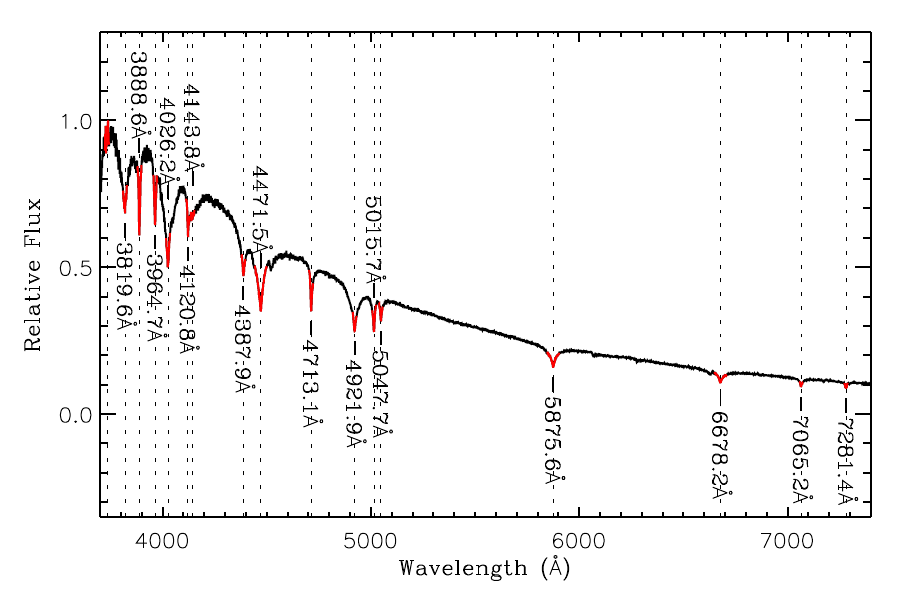}
\caption{Wavelength of the main He {\small \bf I} lines in a DB spectrum.
The positions of the line center, shown as the black dashed lines, are accurate to one decimal place.}
\label{fig:hei}
\end{figure}

DBWDs account for about 20\% of all WDs and have atmospheres dominated by neutral helium, represented as the He {\small \bf I} line in the spectrum.
Compared with all other spectra, the most significant spectral line in a DBWDs spectrum is the He {\small \bf I} at 4471\AA\ \citep{2016MNRAS.455.3413K}. 
The spectral lines shown in Figure \ref{fig:hei} were checked using the atomic line table from the National Institute of Standards and Technology \citep{NIST} to aid the analysis of the feature extraction results. 
In this plot, we show a typical DB spectrum with all of the He {\small \bf I} lines ranging from 3800 to 7400\AA.

Instead of full spectral template matching, ML can be applied to imitate visual recognition to detect the features in noisy spectral data.
The line wings, rather than the line center, may be more sensitive to the distinction between DB and non-DB. The LASSO algorithm has the ability to obtain such positions at some particular wavelength as features, which satisfy the demand. 
In other words, some dominant spectral lines do not work when spectra are similar, such as DB and B that both have a He {\small \bf I} line at 5015.7\AA, which cannot be used as a component of the classifier.

\subsubsection{LASSO}

LASSO is an interpretable model that minimizes the residual sum of matrixes subject to the sum of the absolute value of the coefficients smaller than a constant \citep{1996JRS...58...1}. This model is successfully applied to extract linearly supporting features from stellar spectra, and the atmospheric parameters are automatically estimated including $T_{\text{eff}}$, log $g$, and [Fe/H] \citep{2015ApJS..218....3L}.

In simple terms, consider a sample consisting of $N$ spectra, each of which includes $n$ points. 
Let $y_{i}$ be the outcome and $x_{i}$ be the covariate vector for the $i$th case.
Then, the objective of LASSO is to solve;

\begin{eqnarray*}
\hat{w} = \arg \min_{w}\left\{\sum_{i=1}^{N}\left(y_{i}-f\left(x^{i};w\right)\right)^{2}+\lambda\|w\|_{1}\right\},
\end{eqnarray*}

where 

\begin{eqnarray*}
\|w\|_{1} &=& \sum_{i=1}^{n}|w_{i}|, \\
f\left(x^{i}; w\right) &=& \sum_{i=1}^{n}w_{i}x_{i}
\end{eqnarray*}
and $\lambda > 0$ is a tuning parameter that controls the value of non-zero parameters, $w$, and the complexity of the model.

\cite{2017Springer} has proved that LASSO can effectively filter out most of the irrelevant or redundant variables by reducing the amount of non-zero parameters of $w_{i}$.
We use the LASSO program based on MATLAB \citep{2004math......6456E}, 
in which the parameter $\lambda$ can be equivalently replaced with the number $m$ of non-zero parameters, $w_{i}$.

\subsubsection{Feature Selection}
Here, we adopt LASSO to extract features between the DBWD and other types of spectra.
At first, we build five groups of negative samples for each CPS with the highest S/N\_g (see Table \ref{tab:type} for detail), as the features can be affected by data quality or parameters.
Each group has 300 negative samples, and we combine all features within a CPS into one as the final output.
Features from different wavelength ranges may vary, and most DB spectral lines are, in theory, mainly on the blue band of a spectrum.
Thus, positions of the features within 3900 -- 5900\AA\ and 3900 -- 8900\AA\ need to be analyzed.

\begin{enumerate}
\item Feature with different CPSs \\
Features from different CPSs are not similar, which represent the difference between spectra of this type and DB spectra.
Figure \ref{lasso:diffclass} is one example that illustrates distinct features extracted from WD and QSO groups, which are shown by the short solid blue and red lines, respectively.
From the experiment, we note that the features extracted from QSO almost cover all wavelengths, indicating that the difference between QSO and DB is large.
This is also the main reason why the majority of QSO spectra with high confidence are excluded from the ED in advance.
Conversely, features from the CPSs of A star groups are mainly near 4026.2, 4471.5, 4713.1, 4921.9, 5875.6, and 4341.7\AA, which is the position of the He {\small \bf I} and H$\gamma$ lines.
Face that none of the He {\small \bf I} line exist in normal early A stars, as panel (a) of Figure 2 in \cite{2007PASJ...59..245T} shows, these locations represent the major differences between A and DB stars.

\begin{figure}
\centering
\includegraphics[width=0.49\textwidth]{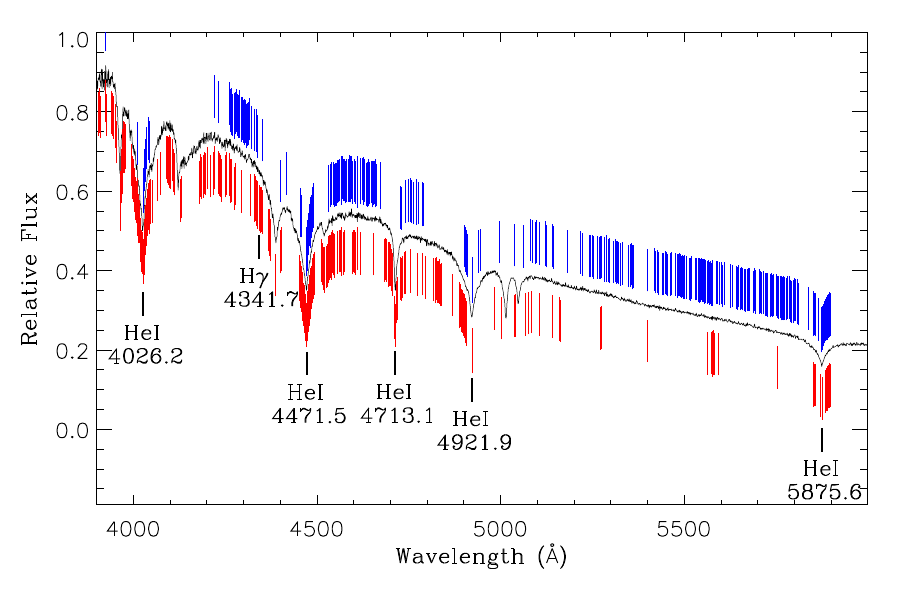}
\caption{Different features from the A and QSO groups with a wave range of from 3900 to 5900\AA.
The features that represent the distinction between A and DB are plotted in the red lines (below), and that of QSO and DB in blues lines (above).}
\label{lasso:diffclass}
\end{figure}

\begin{figure}
\centering
\includegraphics[width=0.48\textwidth]{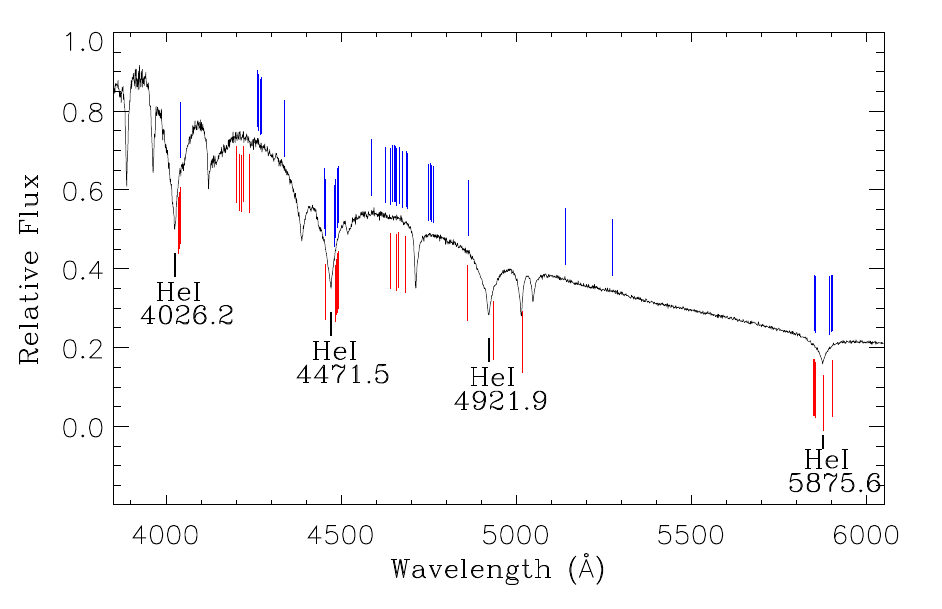}
\caption{Features between different groups with identical type.
The blue and red short lines indicate the wavelength of the feature detected by two data groups separately.
Most of positions within these two groups are similar.}
\label{fig:diffgroup}
\end{figure}

There are five groups in CPS in which negative samples are all classified as O or OB in the DR14 catalog. 
Figure \ref{fig:diffgroup} shows the features extracted from two groups.
It can be shown that the main features of the same classification are similar, but the details are slightly different.
All these spectra were observed by the telescope on Earth, which may lead to uncertainties in the data due to noise from sky light or instrument efficiencies.
Besides, ML is a data-based approach, providing practical results that differ from the results obtained from the theory of spectral analysis.
Therefore, a few more wavelengths, that are not characteristic spectral lines, are recognized as features between O and DB.
We show these features in Table \ref{tab:finalfeature2} in Section \ref{sec:conclusion}.\\

\item Feature within different wavelength range \\
The characteristics in the blue bands (3900 -- 5900\AA) are not exactly the same as those in full spectrum (3900 -- 8900\AA). 
The possible cause is changes in the original points, while the wavelength increases.
Figure \ref{fig:diffwave}, for instance, compares features within 3900 -- 5900\AA (blue, above) and 3900 -- 8900\AA\ (red, below).

\begin{figure}
\centering
\includegraphics[width=0.48\textwidth]{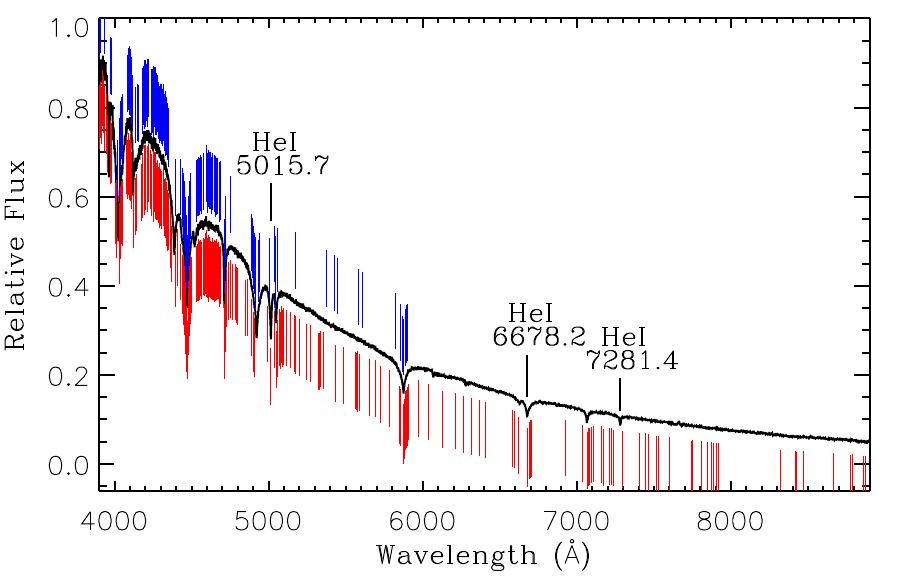}
\caption{Features between B and DB in different wavebands.
The blue lines above the spectrum represent features extracted from 3900 to 5900\AA, while the red lines are those from 3900 to 8900\AA.}
\label{fig:diffwave}
\end{figure}

In Figure \ref{fig:diffwave}, all features within the blue band of the two groups are almost the same, except for some minor differences.
For example, He {\small \bf I} 5015.7\AA\ is an important feature within the wavelength range 3900 -- 5900\AA, but it is not so significant within 3900 -- 8900\AA.
In addition, many data points, such as 7010\AA, are also significant characteristics, that cannot be ignored.
Despite the importance of the blue band in early type stars and DBWDs, there obviously exists features for wavelengths redder than 5900\AA, 
such as H$\alpha$ (6564.6\AA), He {\small \bf I} (6678.2, 7065.2, and 7281.4\AA), and other positions that theoretically have no spectral lines in low resolution spectra.

In conclusion, we use the full spectrum to extract features when performing the following experiment.

\end{enumerate}

\subsection{Features at Multi-scale}

Wavelet decomposition is adopted to analyze features of DBWD at multi-scales. 

\cite{2015ApJS..218....3L} evaluated the performance of various wavelet basis functions and decomposition levels for the estimation of stellar atmospheric parameters, using several evaluation methods. In some situations, the most essential difference between wavelet applications is the selection of the basis function and decomposition level. The efficiency can differ when some variables change, such as the basis function and decomposition level. For example, Meyer and Biorthogonal wavelets can sometimes lead to a large distinction when estimating $T_{\text{eff}}$ of some specific stellar spectra (Panel (e) of Figure 7 in \cite{2015ApJS..218....3L}).

When it comes to analyzing the distribution of DBWD features extracted by LASSO, the wavelet basis function becomes less important since the classification algorithm is only concerned with the wavelength location of the features.
We compare different basis functions and obtain almost similar locations of the features, which is relatively fixed at some wavelengths in the fourth or fifth wavelet coefficients. As a result, we simply employ the simplest basis form --- Haar wavelet --- to conduct the decompose procedure.

Some of the most important features stay at the same position of line wings on the same scale for DBWDs with various temperatures and gravities. Features extracted from one group are served as an example shown in Figure \ref{fig:diffgroup}, in which line wings of He {\small \bf I} (4026.2, 4471.5, 4921.9, and 5875.6\AA), instead of line centers, are recognized as features. 

\begin{figure}
\centering
\includegraphics[width=0.48\textwidth]{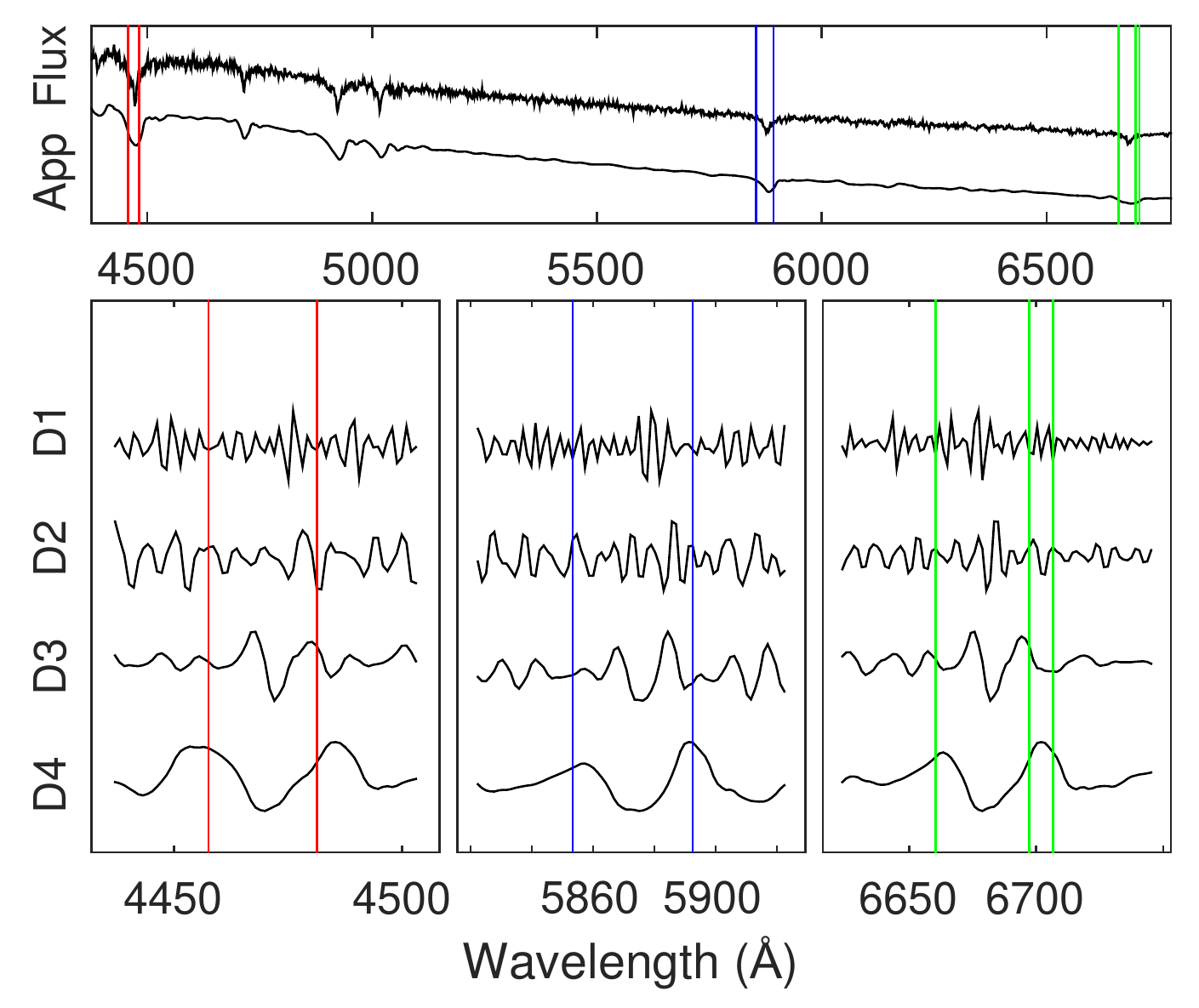}
\caption{Wavelet decomposition of a DB spectrum.
In the upper panel, the data from top to bottom are the original flux of the spectrum and the approximation coefficients after the fourth level decomposition.
The detail coefficients at the first, second, third, and fourth levels are shown in the three bottom panels.}
\label{fig:decomp}
\end{figure}

We decompose a spectrum into a low-frequency approximation signal and high-frequency details by wavelet transform, and discover that most features fall on the crest of the detail coefficients at the fourth layer of the wavelet domain.
For example, 4471.5, 5875.6, and 6678.2\AA\ in Figure \ref{fig:decomp} are the three main typical features in DB.

The features located at the 16th point from the line center or nearby, coincide with the coefficients of the fourth layer of the wavelet decomposition ($2^{4}$).
This part of the spectral line should become a major character of WDs when distinguishing from other types of spectra.
We believe that spectral line broadening is a significant characteristic of WDs and that the positions are the most dramatic changes in the spectral data.

\subsection{SVM}
\label{sec:trainsvm}

\subsubsection{Hyper-plane}

As a supervised ML method, SVM is used in classification and regression analysis.
In a dual clustering system, given a set of training examples that are each marked by positive or negative categories, an SVM training algorithm builds a robust binary linear classifier model that sorts new data to one type or the other. 

We apply the LIBSVM \citep{CC01a} software to pick DB spectra from all of the data set with features obtained in Section \ref{sec:feature}.
LIBSVM is an integrated software for support vector classification, regression, and distribution estimation.
It supports multi-type classifications.

Let $(x_{i},y_{i}), i = 1,2,\ldots,n$ represent a training set, 
where $x_{i} \in R^{n}$ and $y \in \{-1,1\}^{n}$ are the spectral data at the feature points and label, respectively.
LIBSVM tries to seek a linear separating hyper-plane with the maximal margin in this higher dimensional space by solving the following optimization problem, 

\begin{eqnarray*}
\min_{w,b,\xi} & \frac{1}{2}w^{T}w + C\sum_{i=1}^{n}\xi _{i} \nonumber \\
\text{subject to} & y_{i}\left(w^{T}\phi(x_{i})+b\right)\ge 1-\xi _{i}, \\
 & \xi _{i} \ge 0. \nonumber
\end{eqnarray*}
Here training vectors $x_{i}$ are mapped into a higher dimensional space by the function $\phi$ and $C>0$ is the penalty parameter of the error term.
Furthermore, $K(x_{i}, x_{j}) = \phi (x_{i})^{T} \phi(x_{j})$ is the kernel function.
LIBSVM provides four basic kernels below:
\begin{enumerate}
\item linear: $K(x_{i}, x_{j}) = x_{i}^{T}x_{j}$.
\item polynomial: $K(x_{i}, x_{j}) = (\gamma x_{i}^{T}x_{j}+r)^{d}, \gamma >0$.
\item radial basis function (RBF): $K(x_{i}, x_{j}) = \text{exp}(-\gamma\|x_{i}-x_{j}\|^{2}),\gamma > 0$.
\item sigmoid: $K(x_{i}, x_{j}) = \text{tanh}(\gamma x_{i}^{T}x_{j}+r)$.
\end{enumerate}
Here, $\gamma$, $r$, and $d$ are the kernel parameters.

The kernel function and parameters are important for the SVM algorithm to be adjusted.
Experiments show that the linear and RBF kernels should provide better discrimination results for spectral data.
10-fold cross-validation is utilized to automatically determine all parameters using LIBSVM software.

\subsubsection{Verification}
\label{sec:verification}

In this section, we verify the reliability of the algorithm by labeling all of the testing set.
There are some measures for information retrieval and statistical classification to evaluate the quality of the algorithm: accuracy, precision, and recall.
The accuracy is based on our prediction, which shows how many of the positive predictions are true positives (TPs).
The recall rate shows how many positive examples in the sample were predicted correctly.
We use the following terms: TP for the correct prediction of the positive category; 
false positive (FP) for that of the incorrect positive category; and 
false negative (FN) and true negative (TN) for incorrect and TNs, respectively.
During the training process, almost all of the positive samples, except a few, have been recognized correctly. 
Therefore, the recall TP/(TP+FN) can reach an approximate percentage of 100\%.
Generally, the mean accuracy (TP+TN)/(TP+FN+FP+TN), and precision TP/(TP+FP), of all of the CPS can reach 99.9\% and 99.7\%, respectively, 
indicating a very high stability and reliability for this algorithm.

\section{Recognition}
\label{sec:recognition}
\subsection{Input of the SVM}

\begin{figure*}
\centering
\includegraphics[width=0.98\textwidth]{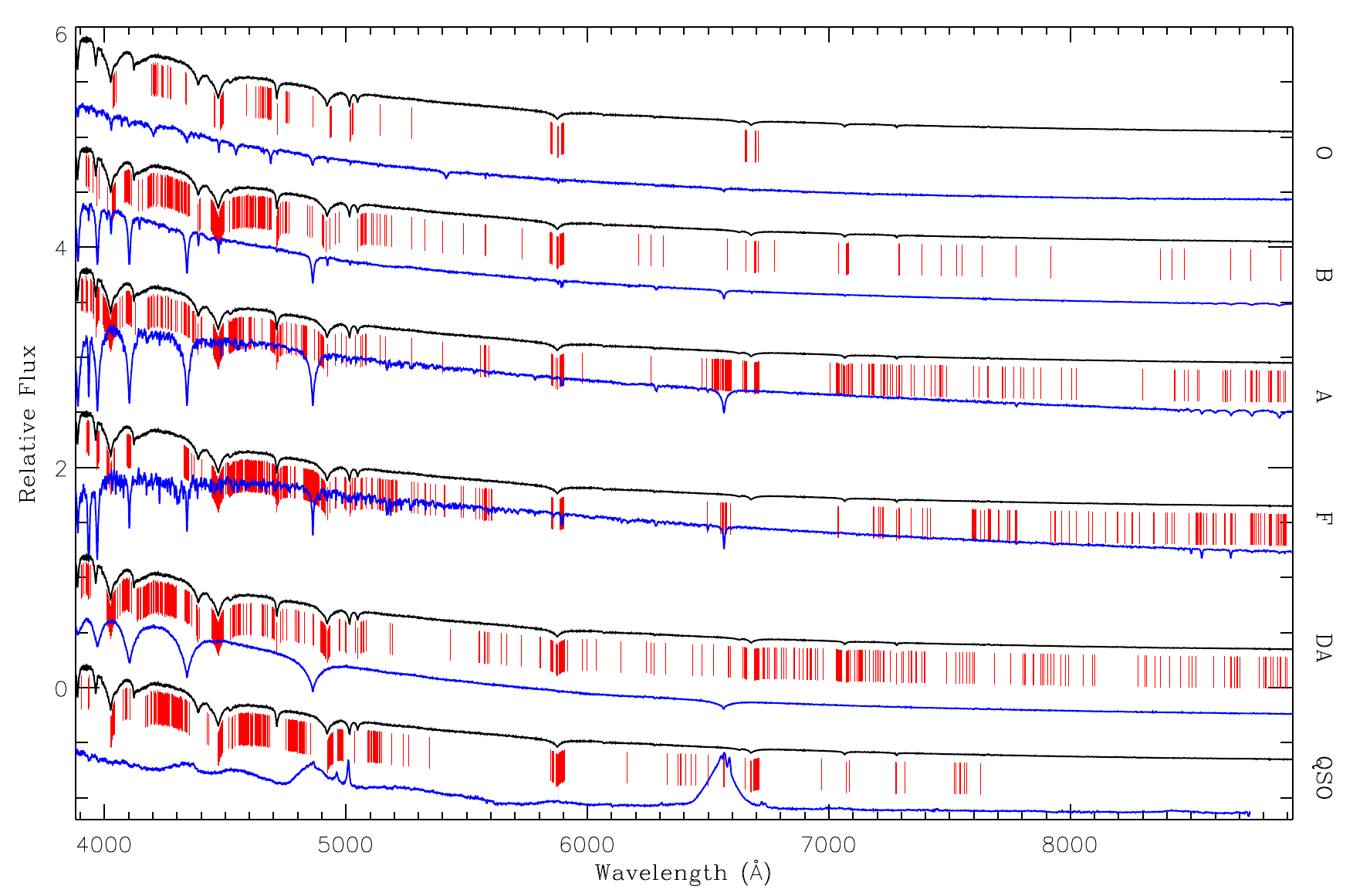}
\caption{Features of DB versus O, B, A, F, WD(DA), and QSO, from top to bottom in order.
To make it more explicit, the wavelength of the features are marked with red lines below each DB spectrum (black); the other six types of spectra are plotted in blue.}
\label{fig:finalfeature}
\end{figure*}

We collect all of the features derived from Section \ref{sec:feature}, and demonstrate some of them obtained from the CPSs of O, B, A, F, WD, and QSO (from top to bottom) in Figure \ref{fig:finalfeature}.

The features are marked with short red lines between the DB spectra and that of the other CPSs. 
In general, features on either side of the spectral lines are not perfectly symmetrical.
Distinctions near He {\small \bf I} 4026.2\AA, for instance, between DB and O, only appeared on the right side of this He {\small \bf I} line.
It can be intuitively derived that a large number of features are either just a single data points or narrow ranges of wavelengths.
When they exist in a relatively long range of wavelength, this part should display the most dramatic changes in the spectral lines, as discussed below.
The content of metal elements increases with changes in the stellar types O, B, A, and F, 
corresponding to a raise in the number of features within the red band.

\begin{figure}
\centering
\includegraphics[width=0.35\textwidth]{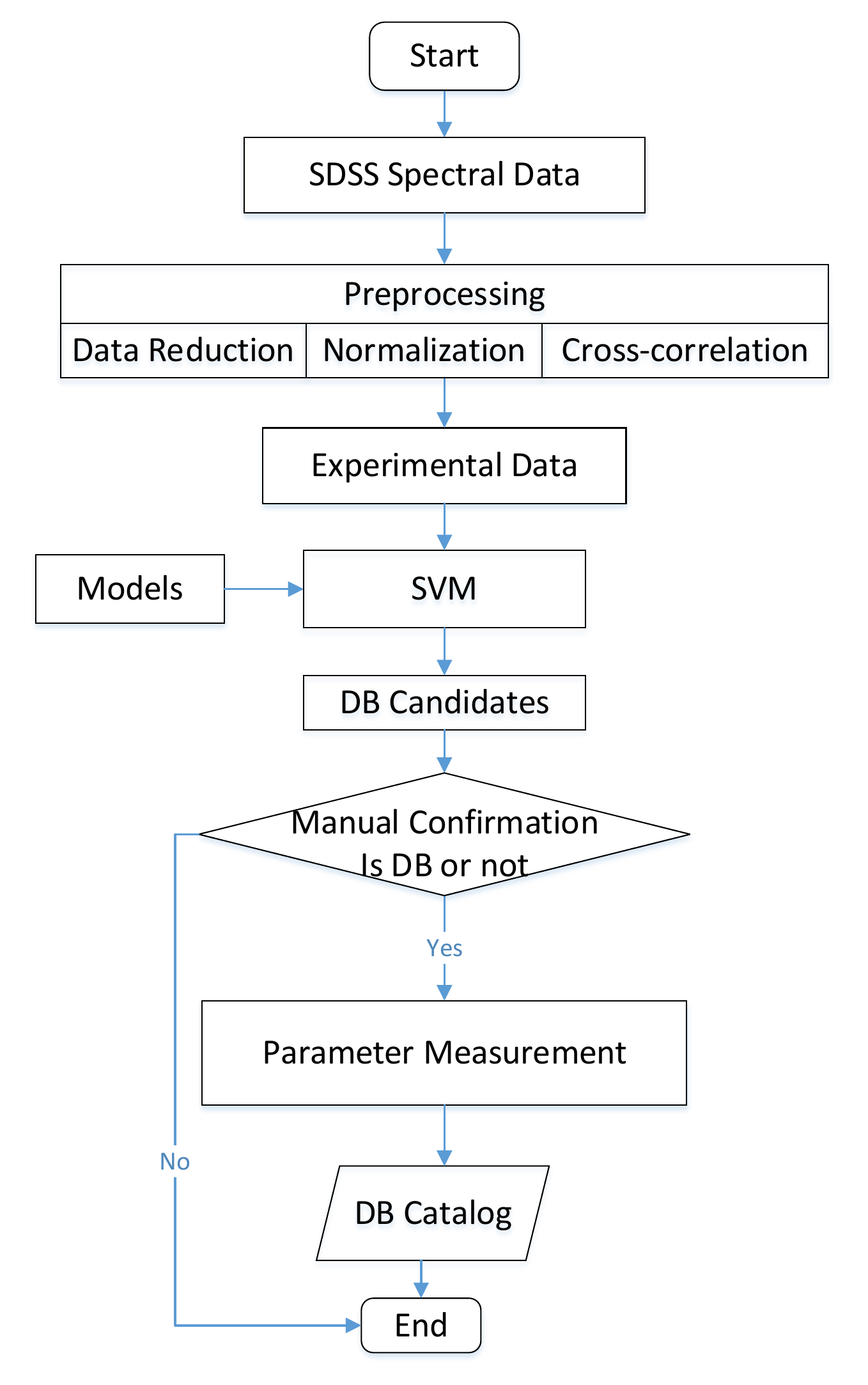}
\caption{Flowchart of the recognition procedure.
This is the application stage of Figure \ref{flch:all} and Section \ref{sec:method}.
The final catalog will also be generated in this part.}
\label{flch:libsvm}
\end{figure}

Afterwards, the DB candidates will be identified from all of the spectra during the searching process. 
The models of the feature derived from the training process, Section \ref{sec:feature}, are employed as the input of SVM.
A flowchart is presented in Figure \ref{flch:libsvm} to demonstrate the start-to-end flow of this procedure. 

\subsection{Recognition and Results}

After the reduction procedure described in Section \ref{sec:datareco}, the ED is generated as the original data set to conduct recognition.
Similar to the prepocessing in Section \ref{sec:preprocess}, we normalize the ED and then move all of them to the rest frame using cross-correlation. 

Then a hyper-plane in feature space is applied by SVM to distinguish the DBWDs from the ED.
We inspect the output and find a sample of 2808 spectra of 2029 different objects from SDSS DR12 and DR14, in which 58 are newly identified objects.
In Table \ref{tab:result}, we illustrate those spectra classified as DBWDs to evaluate the performance of the algorithm model.
In this part, we could not provide the precise ratio mentioned above because we cannot confirm how many real DBWDs reside in the predication negative category.
Assuming all labeled negative samples are non-DB, then the mean percent of correctly identified samples using the algorithm can reach 99.5\%.

\begin{table}
\caption{Results of the experiment and evaluation of the algorithm model.}
\label{tab:result}
\centering
\begin{threeparttable}
\begin{tabular}{ccrrrr} \hline
Class\tnote{a}	&	Subclass\tnote{a}	&	Numbers in ED\tnote{b}	&	DB Candidate\tnote{c}	&	DB\tnote{d}	&	Ratio\tnote{e}	\\	\hline
Star	&	O			&	6497		&	1569	&	1522	&	99.9\%	\\
Star	&	B			&	14,759		&	324		&	115		&	98.6\%	\\
Star	&	A			&	85,468		&	82		&	10		&	99.7\%	\\
Star	&	F			&	192,387		&	1750	&	14		&	99.1\%	\\
Star	&	G			&	101,230		&	62		&	2		&	98.9\%	\\
Star	&	K			&	79,775		&	559		&	3		&	99.3\%	\\
Star	&	M			&	69,675		&	181		&	101		&	99.7\%	\\
Star	&	L			&	5678		&	19		&	7		&	99.9\%	\\
Star	&	T			&	1676		&	9		&	6		&	99.9\%	\\
Star	&	WD			&	31,776		&	423		&	64		&	98.9\%	\\
Star	&	CV			&	9788		&	248		&	19		&	97.6\%	\\
Star	&	carbon		&	3088		&	26		&	2		&	99.2\%	\\
Galaxy	&	broadline	&	6587		&	113		&	1		&	98.3\%	\\
Galaxy	&	null		&	192,319		&	198		&	34		&	99.9\%	\\
QSO		&	broadline	&	97,867		&	237		&	12		&	99.4\%	\\
QSO		&	null		&	87,980		&	1264	&	862		&	99.4\%	\\	\hline
\multicolumn{2}{c}{Total}	&	883,644	&	6952	&	2774	&	99.5\%	\\ \hline
\end{tabular}
\begin{tablenotes}
\item[a] ``Class'' and ``subclass'' are adopted from the data archive of SDSS.
\item[b] Number of spectra of every CPS in the ED.
\item[c] Number of positive samples in every CPS directly derived from the SVM.
\item[d] Number of positive samples in every CPS after visual inspection.
\item[e] Approximation of the identification precision when the predication negative samples are all correct, i.e., the correct proportion of classifications.
\end{tablenotes}
\end{threeparttable}
\end{table}

Clearly, most DBWDs are identified from O, B, WD ,and QSO in the DR14 data set.
In a different way, this also indicates that the qualities, such as spectral lines, of DBWDs are more likely with these types;
or perhaps they are usually mixed together when matched with the full spectrum instead of some particular wavelengths (or features).

\begin{table}
\caption{Target Selection of DBWDs in the SDSS DR12 and DR14.}
\label{tab:source}
\centering
\begin{threeparttable}
\begin{tabular}{lrrrr} \hline
Source\tnote{a}			&	QSO\tnote{b}	&	Galaxy\tnote{b}	&	Star\tnote{b}	&	Total	\\ \hline
AMC						&	1				&	1				&	---				&	2		\\
ELG						&	---				&	---				&	1				&	1		\\
HOT\_STD				&	22				&	3				&	514				&	539		\\
LRG						&	2				&	---				&	1				&	3		\\
NONLEGACY				&	24				&	3				&	413				&	440		\\
Null					&	4				&	---				&	3				&	7		\\
QA						&	---				&	---				&	2				&	2		\\
QSO						&	1				&	---				&	20				&	21		\\
QSO\_EBOSS\_W3\_ADM		&	1				&	---				&	2				&	3		\\
QSO\_VAR				&	---				&	---				&	1				&	1		\\
QSO\_VAR\_SDSS			&	1				&	---				&	1				&	2		\\
ROSAT\_D				&	---				&	---				&	2				&	2		\\
SEGUE1					&	5				&	---				&	1				&	6		\\
SEQUELS\_TARGET			&	1				&	1				&	3				&	5		\\
SERENDIPITY\_BLUE		&	29				&	5				&	180				&	214		\\
SERENDIPITY\_DISTANT	&	23				&	3				&	207				&	233		\\
STAR					&	---				&	---				&	2				&	2		\\
STAR\_CATY\_VAR			&	1				&	---				&	11				&	12		\\
STAR\_WHITE\_DWARF		&	6				&	3				&	84				&	93		\\
WHITEDWARF\_NEW			&	462				&	3				&	236				&	701		\\
WHITEDWARF\_SDSS		&	295				&	2				&	188				&	485		\\\hline
Total					&	878				&	24				&	1872			&	2774	\\ \hline
\end{tabular}
\begin{tablenotes}
\item[a] Target selection of SDSS DR12 and DR14.
\item[b] ``Class'' and ``subclass'' are adopted from the data archive of SDSS.
\end{tablenotes}
\end{threeparttable}
\end{table}

The target selections of all DBWDs are given in Table \ref{tab:source}.
Many spectra, with sources that are stars, are mis-classified into QSO, such as ``WHITEDWARF\_NEW'' listed in Table \ref{tab:source}.
We believe this is due to a weakness in the algorithm of pipeline. 
Without an efficient feature wavelength, many spectra may not be correctly classified by full spectral template matching.
There are also some quite broad spectral lines in both the QSO and WD (DB) spectra that may mislead the classification results.

\section{Analysis}
\label{sec:analysis}

\subsection{Literature Comparison}

Altogether there are 1309 pure DB objects, including double stars, in the literature (see Section \ref{sec:idlit}).
In this paper, we present 1999 objects of DBWDs (including, but not limited to, DB, DBA, or DBZ) with 2774 spectra in SDSS DR12 and DR14, among which 58 objects are newly spectroscopically confirmed.

\begin{longrotatetable}
\begin{deluxetable*}{cccccrrrrrrrl}
\tablecaption{Newly spectroscopically confirmed DBWDs from SDSS DR12 and DR14.\label{final:catalog}}
\tablewidth{700pt}
\tabletypesize{\scriptsize}
\tablehead{
\colhead{Designation} & \colhead{P--M--F} & \colhead{Type} & 
\colhead{RV$_{\text{DB}}$} & \colhead{RV$_{\text{M}}$} & \colhead{$T_{\text{eff}}$} & \colhead{log $g$} & 
\colhead{FUV} & \colhead{NUV} & \colhead{S/N} & \colhead{Mass} & \colhead{Age} & \colhead{Ref} \\ 
\colhead{} & \colhead{} & \colhead{} & 
\colhead{(km s$^{-1}$)} & \colhead{(km s$^{-1}$)} & \colhead{(K)} & \colhead{(cgs)} &
\colhead{(mag)} & \colhead{(mag)} & \colhead{} & \colhead{M$_\odot$} & \colhead{Myr} & \colhead{}
} 

\startdata
J094038.80+364645.6 & 1275--52996--0037 & DB & 449$\pm$0 & \nodata & 18200$\pm$2166 & 8.09$\pm$0.223 & 19.8$\pm$0.2 & 19.1$\pm$0.1 & 2.5 & 0.60 & 106.0 & 0 \\
J115601.31+293115.4 & 2224--53815--0171 & DB & -41$\pm$31 & \nodata & 42897$\pm$1577 & 7.51$\pm$0.088 & 19.5$\pm$0.1 & 19.7$\pm$0.1 & 3.2 & -9999 & -9999 & 0 \\
J000801.20+272906.1 & 2824--54452--0037 & DBAZ & 149$\pm$32 & \nodata & 16673$\pm$757 & 7.92$\pm$0.225 & \nodata & \nodata & 4.7 & 0.59 & 130.7 & 0 \\
J222646.14+061921.3 & 4410--56187--0506 & DB & 70$\pm$37 & \nodata & 20120$\pm$1567 & 8.75$\pm$0.159 & 20.0$\pm$0.1 & 19.8$\pm$0.1 & 10.8 & 0.91 & 204.3 & 0 \\
J012752.18+140622.9 & 4665--56209--0726 & DBA & -28$\pm$16 & \nodata & 32695$\pm$441 & 8.87$\pm$0.023 & \nodata & \nodata & 22.1 & 1.20 & 148.2 & 0 \\
J222711.11+073510.7 & 5057--56209--0276 & DB & 50$\pm$12 & \nodata & 15391$\pm$82 & 8.88$\pm$0.029 & 19.7$\pm$0.1 & 18.9$\pm$0.1 & 22.7 & 1.19 & 1100.0 & 0 \\
J095403.47+223919.2 & 5787--56254--0254 & DB & 118$\pm$59 & \nodata & 13661$\pm$339 & 8.17$\pm$0.111 & 20.5$\pm$0.2 & 19.5$\pm$0.1 & 13.2 & 0.59 & 245.9 & 0 \\
J094852.66+233004.1 & 5787--56254--0500 & DB & 58$\pm$14 & \nodata & 17018$\pm$157 & 8.75$\pm$0.044 & 18.8$\pm$0.1 & 18.4$\pm$0.1 & 11.1 & 1.19 & 822.0 & 0 \\
J094852.66+233004.1 & 5788--56255--0028 & DB & 68$\pm$12 & \nodata & 17615$\pm$129 & 8.77$\pm$0.030 & 18.8$\pm$0.1 & 18.4$\pm$0.1 & 16.7 & 1.19 & 705.9 & 0 \\
J090730.35+270413.6 & 5780--56274--0018 & DB & 113$\pm$22 & \nodata & 18200$\pm$275 & 8.88$\pm$0.048 & 19.8$\pm$0.1 & 19.3$\pm$0.0 & 10.1 & 1.19 & 705.9 & 0 \\
J100104.94+302543.5 & 5800--56279--0890 & DB & 27$\pm$33 & \nodata & 13939$\pm$110 & 8.92$\pm$0.053 & 20.2$\pm$0.2 & 19.3$\pm$0.1 & 12.2 & 1.19 & 1268.0 & 0 \\
J135815.93+290525.5 & 6009--56313--0624 & DB & 46$\pm$3 & \nodata & 22148$\pm$208 & 8.76$\pm$0.013 & 17.1$\pm$0.1 & 16.9$\pm$0.0 & 30.1 & 1.19 & 372.9 & 11 \\
J091256.90+430023.0 & 4687--56338--0324 & DB & 4$\pm$24 & \nodata & 13893$\pm$184 & 9.03$\pm$0.073 & \nodata & \nodata & 11.1 & 1.19 & 1268.0 & 0 \\
J091256.90+430023.0 & 4687--56369--0326 & DB+M1 & 181$\pm$2893 & 1147$\pm$52 & 17954$\pm$409 & 9.32$\pm$0.087 & \nodata & \nodata & 11.8 & -9999 & -9999 & 0 \\
J091638.24+475253.6 & 5813--56363--0640 & DB & -7$\pm$7 & \nodata & 17900$\pm$99 & 8.67$\pm$0.024 & 18.4$\pm$0.1 & 18.0$\pm$0.0 & 17.5 & 0.91 & 278.0 & 0 \\
J142046.13+554201.4 & 6803--56402--0201 & DB & 217$\pm$60 & \nodata & 17200$\pm$778 & 8.34$\pm$0.179 & \nodata & \nodata & 3.0 & 0.91 & 326.3 & 0 \\
J091534.70+513610.3 & 5729--56598--0121 & DB & 37$\pm$17 & \nodata & 16071$\pm$115 & 8.77$\pm$0.041 & 19.4$\pm$0.2 & 19.0$\pm$0.1 & 12.1 & 1.19 & 951.5 & 0 \\
J092540.36+511229.6 & 5730--56607--0940 & DB & 45$\pm$30 & \nodata & 17247$\pm$293 & 8.78$\pm$0.049 & \nodata & \nodata & 9.6 & 1.19 & 822.0 & 0 \\
J012644.96-025633.9 & 7877--56898--0048 & DB & 172$\pm$2893 & \nodata & 13162$\pm$949 & 8.89$\pm$0.460 & \nodata & \nodata & 2.6 & 1.19 & 1458.0 & 0 \\
J080710.33+485259.6 & 7324--56935--0828 & AMCVn & 462$\pm$56 & \nodata & 8260$\pm$23 & 8.18$\pm$0.032 & \nodata & \nodata & 3.5 & -9999 & -9999 & 0 \\
J231213.74+185713.8 & 7611--56946--0897 & DBO & 1260$\pm$593 & \nodata & 26200$\pm$18432 & 6.33$\pm$1.958 & \nodata & \nodata & 1.9 & -9999 & -9999 & 0 \\
J022756.30--044504.2 & 8127--56957--0899 & DB & 1245$\pm$107 & \nodata & 7653$\pm$7143 & 7.27$\pm$38.056 & \nodata & \nodata & 0.7 & -9999 & -9999 & 0 \\
J012920.84+191241.5 & 7628--56978--0465 & DB & 877$\pm$48 & \nodata & 9228$\pm$221 & 7.23$\pm$0.141 & \nodata & \nodata & 0.8 & 0.19 & 351.9 & 0 \\
J235607.30+025254.8 & 7849--56980--0914 & DBZ & 99$\pm$38 & \nodata & 16037$\pm$236 & 8.86$\pm$0.087 & 20.5$\pm$0.2 & 20.0$\pm$0.1 & 6.8 & 1.19 & 951.5 & 0 \\
J020022.48+242343.3 & 7692--57064--0409 & DB & 1245$\pm$1076 & \nodata & 18200$\pm$6743 & 6.08$\pm$3.551 & \nodata & \nodata & -0.0 & -9999 & -9999 & 0 \\
J074325.35+432027.7 & 8276--57067--0470 & DB & 149$\pm$44 & \nodata & 13841$\pm$1083 & 8.88$\pm$0.500 & \nodata & \nodata & 2.7 & 1.19 & 1268.0 & 0 \\
J005436.04--041940.6 & 7912--57310--0460 & DBA & 268$\pm$28 & \nodata & 16939$\pm$508 & 6.58$\pm$0.162 & 18.9$\pm$0.1 & 19.4$\pm$0.1 & 5.5 & -9999 & -9999 & 0 \\
J013634.37--001109.9 & 8792--57364--0358 & DBA & 27$\pm$58 & \nodata & 16415$\pm$200 & 8.88$\pm$0.056 & 20.3$\pm$0.2 & 20.0$\pm$0.1 & 9.4 & 1.19 & 951.5 & 0 \\
J234924.30--025209.5 & 7851--56932--0403 & DB & 1152$\pm$70 & \nodata & 17200$\pm$4598 & 6.19$\pm$1.638 & \nodata & \nodata & 1.6 & -9999 & -9999 & 0 \\
J232933.23+212015.6 & 7604--56947--0864 & DB & 1127$\pm$77 & \nodata & 18273$\pm$1310 & 8.95$\pm$0.137 & \nodata & \nodata & 2.2 & 1.19 & 705.9 & 0 \\
J082623.07+555006.2 & 7375--56981--0144 & DB & -299$\pm$87 & \nodata & 15200$\pm$1122 & 9.46$\pm$0.366 & \nodata & \nodata & 6.7 & -9999 & -9999 & 0 \\
J081453.04+555033.1 & 7375--56981--0487 & DB & -149$\pm$74 & \nodata & 17200$\pm$738 & 8.95$\pm$0.178 & \nodata & \nodata & 3.2 & 1.19 & 822.0 & 0 \\
J234527.43+215712.3 & 7600--56984--0082 & DB & 1315$\pm$696 & \nodata & 28200$\pm$3881 & 8.77$\pm$0.184 & \nodata & \nodata & 5.2 & 1.19 & 184.3 & 0 \\
J234131.56+224240.6 & 7600--56984--0313 & DB+M9 & 67$\pm$35 & 1174$\pm$30 & 21428$\pm$791 & 8.84$\pm$0.049 & \nodata & \nodata & 9.5 & 1.19 & 436.0 & 0 \\
J025818.61--004131.3 & 7820--56984--0106 & DB & 308$\pm$134 & \nodata & 20158$\pm$1915 & 7.90$\pm$0.203 & \nodata & \nodata & 3.2 & 0.60 & 68.8 & 0 \\
J025720.06--003812.0 & 7820--56984--0182 & DB & 884$\pm$59 & \nodata & 15400$\pm$992 & 7.84$\pm$0.282 & \nodata & \nodata & 2.1 & 0.59 & 198.7 & 0 \\
J225345.44+223258.6 & 7613--56988--0380 & DB+M9 & 29$\pm$37 & 1164$\pm$41 & 12473$\pm$486 & 9.35$\pm$0.190 & 22.0$\pm$0.3 & 21.1$\pm$0.1 & 7.8 & -9999 & -9999 & 0 \\
J093021.79+544359.5 & 7285--56991--1000 & DB & -2098$\pm$81 & \nodata & 17194$\pm$761 & 8.95$\pm$0.222 & \nodata & \nodata & 3.3 & 1.19 & 822.0 & 0 \\
J080128.06+554004.7 & 7281--57007--0548 & DB & 29$\pm$795 & \nodata & 17707$\pm$1093 & 7.66$\pm$0.262 & 21.4$\pm$0.4 & 21.2$\pm$0.3 & 2.0 & 0.36 & 56.2 & 0 \\
J023303.84--022104.8 & 7829--57011--0397 & DB & 1245$\pm$1578 & \nodata & 16429$\pm$1429 & 8.35$\pm$0.493 & \nodata & \nodata & 1.8 & 0.91 & 385.3 & 0 \\
J084704.97+511056.0 & 7303--57013--0675 & DB & 1245$\pm$2586 & \nodata & 15650$\pm$1067 & 8.58$\pm$0.502 & \nodata & \nodata & 1.5 & 0.91 & 385.3 & 0 \\
J104624.26+490908.9 & 7387--57038--1000 & DB & 173$\pm$41 & \nodata & 36204$\pm$1786 & 8.60$\pm$0.099 & 20.7$\pm$0.3 & 20.5$\pm$0.2 & 3.2 & 0.93 & 16.1 & 0 \\
J091854.84+515603.2 & 7289--57039--0049 & DB & -1498$\pm$76 & \nodata & 20200$\pm$3648 & 7.45$\pm$0.617 & \nodata & \nodata & 2.8 & 0.37 & 39.1 & 0 \\
J010633.14+203043.1 & 7624--57039--0927 & DB & 1191$\pm$23 & \nodata & 21763$\pm$1440 & 8.59$\pm$0.111 & \nodata & \nodata & 3.0 & 0.91 & 152.4 & 0 \\
J090702.69+430612.5 & 8282--57041--0092 & DB & 1140$\pm$97 & \nodata & 15436$\pm$1381 & 7.21$\pm$0.535 & \nodata & \nodata & 3.5 & 0.23 & 77.4 & 0 \\
J085309.15+584336.3 & 8197--57064--0537 & DBAZ & 29$\pm$62 & \nodata & 16715$\pm$466 & 9.23$\pm$0.112 & \nodata & \nodata & 9.2 & 1.19 & 822.0 & 0 \\
J103107.17+520854.7 & 8167--57071--0548 & DB & 216$\pm$24 & \nodata & 16535$\pm$531 & 8.76$\pm$0.146 & 21.3$\pm$0.5 & 20.7$\pm$0.2 & 3.1 & 1.19 & 822.0 & 0 \\
J112752.26+565539.5 & 8176--57131--0392 & DB & 449$\pm$98 & \nodata & 17200$\pm$878 & 7.94$\pm$0.206 & \nodata & \nodata & 2.2 & 0.59 & 130.7 & 0 \\
J120156.39+493707.4 & 7423--57135--0578 & DB+M2 & 29$\pm$37 & 1146$\pm$8 & 29415$\pm$3758 & 8.80$\pm$0.168 & 21.8$\pm$0.5 & 20.9$\pm$0.1 & 3.0 & 1.19 & 166.8 & 0 \\
J121027.14+502735.7 & 7423--57135--0833 & DBZ & 172$\pm$37 & \nodata & 15530$\pm$307 & 9.07$\pm$0.096 & 20.4$\pm$0.3 & 19.7$\pm$0.1 & 13.5 & 1.19 & 951.5 & 0 \\
J214544.31+270923.4 & 7641--57307--0622 & DB & 1245$\pm$364 & \nodata & 21503$\pm$2935 & 7.90$\pm$0.210 & \nodata & \nodata & 2.5 & 0.60 & 43.8 & 0 \\
J223318.14+244812.3 & 7654--57330--0204 & DB & 618$\pm$43 & \nodata & 36971$\pm$2568 & 8.74$\pm$0.135 & \nodata & \nodata & 2.0 & 0.93 & 13.6 & 0 \\
J231304.25+265057.9 & 7703--57333--0554 & DB & 659$\pm$43 & \nodata & 14268$\pm$1085 & 6.83$\pm$0.488 & \nodata & \nodata & 1.7 & 0.22 & 96.8 & 0 \\
J020910.57--043943.1 & 7885--57336--0410 & DB & 1245$\pm$1704 & \nodata & 35349$\pm$25761 & 7.20$\pm$1.613 & \nodata & \nodata & 0.1 & 0.32 & 10.5 & 0 \\
J005242.64+285411.0 & 7674--57359--0834 & DB & 570$\pm$45 & \nodata & 18200$\pm$840 & 9.19$\pm$0.180 & 21.6$\pm$0.4 & 21.1$\pm$0.3 & 3.1 & 1.19 & 705.9 & 0 \\
J001334.89+264245.4 & 7694--57359--0405 & DB & -599$\pm$23 & \nodata & 16025$\pm$343 & 9.44$\pm$0.161 & \nodata & \nodata & 4.3 & -9999 & -9999 & 0 \\
J001627.53+281843.5 & 7694--57359--0737 & DB & 703$\pm$66 & \nodata & 16495$\pm$606 & 7.90$\pm$0.165 & \nodata & \nodata & 2.6 & 0.59 & 161.0 & 0 \\
J075925.84+414454.4 & 8291--57391--0933 & DB+M9 & 29$\pm$68 & 1169$\pm$61 & 16303$\pm$473 & 8.23$\pm$0.108 & 18.8$\pm$0.1 & 19.5$\pm$0.1 & 8.4 & 0.59 & 161.0 & 0 \\
J014803.85+005317.7 & 8793--57391--0826 & DB & 29$\pm$55 & \nodata & 16485$\pm$510 & 8.85$\pm$0.143 & 22.1$\pm$0.4 & 21.7$\pm$0.2 & 2.3 & 1.19 & 951.5 & 0 \\
J131316.23+511428.8 & 8210--57426--0532 & DBA & -2$\pm$41 & \nodata & 14982$\pm$554 & 8.61$\pm$0.261 & \nodata & \nodata & 2.0 & 0.91 & 459.1 & 0 \\
\enddata
\tablecomments{The full table is available online.}
\end{deluxetable*}
\end{longrotatetable}

\begin{table*}
\caption{Columns provided in Table \ref{final:catalog} and online table.}
\label{tab:etable}
\centering
\begin{tabular}{cll} \hline
Column No.	&	Heading				&	Description	\\ \hline
1			&	Designation			&	SDSS object name (SDSS 2000J+)	\\
2			&	P--M--F				&	SDSS Plate number--Modified Julian date--Fiber	\\
3			&	Type				&	Classification of objects derived from ML method	\\
4			&	RV$_{\rm DB}$		&	Radial velocity and uncertainty of each spectrum (km s$^{-1}$)	\\
5			&	RV$_{\rm M}$		&	Radial velocity and uncertainty of M companions (km s$^{-1}$)	\\
6			&	$T_{\text{eff}}$	&	Effective temperature (K)	\\
7			&	log $g$				&	Surface gravity (cgs)	\\
8			&	FUV					&	Magnitude of FUV from GALEX, -9999: there is no corresponding value (mag)	\\
9			&	NUV					&	Magnitude of NUV from GALEX, -9999: there is no corresponding value (mag)	\\
10			&	S/N					&	Median S/N from catalog of SDSS DR14	\\
11			&	Mass				&	Obtained from \cite{coolingmodels} ($M_\odot$)	\\
12			&	Age					&	Obtained from \cite{coolingmodels} (Myr)	\\
13			&	Ref					&	ID of literature, 0: newly identified in this paper, see Section \ref{sec:idlit} for detail \\
\hline
\end{tabular}
\end{table*}

\begin{figure}
\centering
\includegraphics[width=0.8\textwidth]{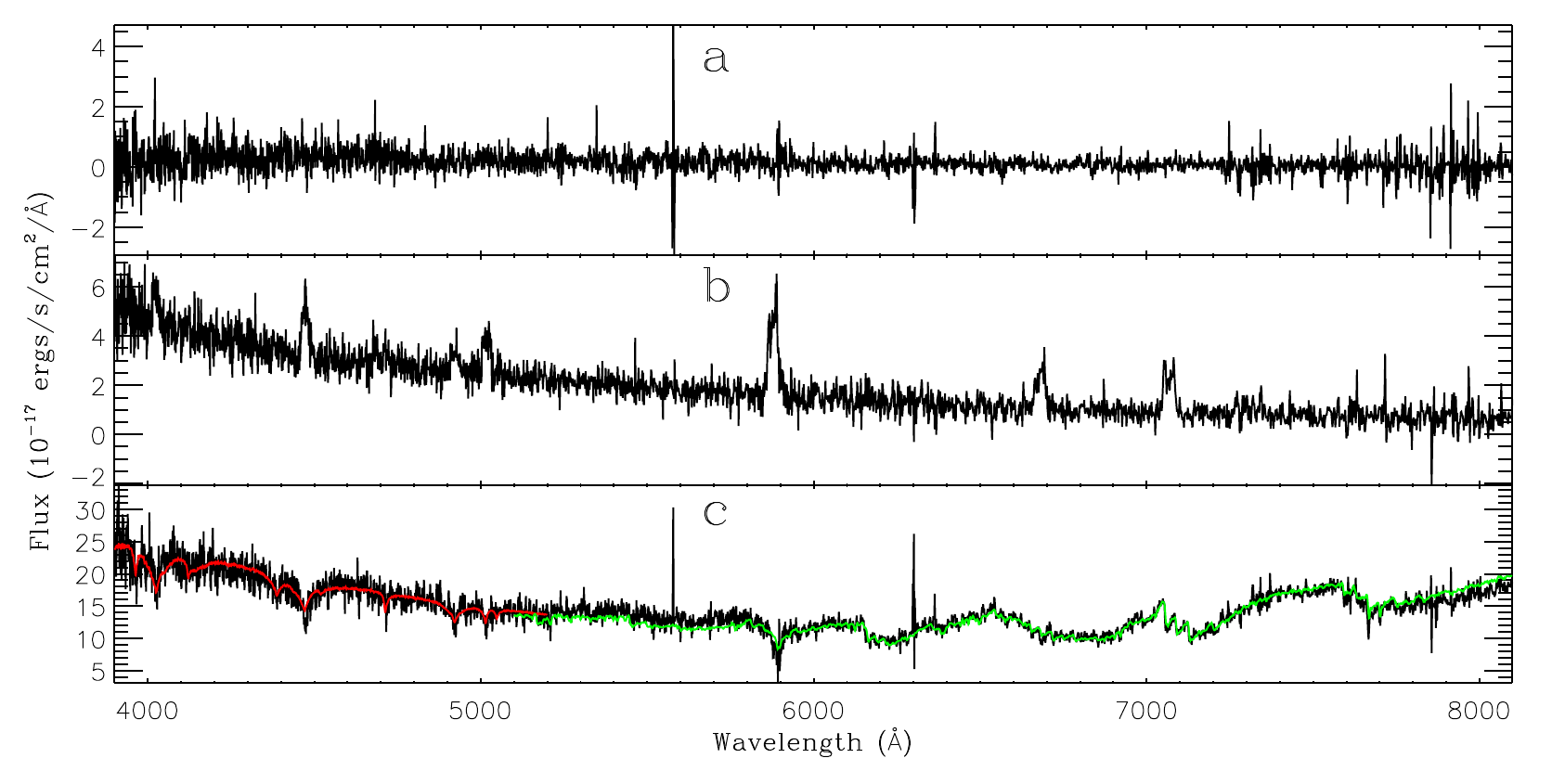}
\caption{Three typical spectra.
The spectrum in panel ``a'' is a DBWD in the literature but not in our catalog.
Panel ``b'' is a newly spectroscopically confirmed AM CVn star.
One of the DB plus M double stars is presented in panel ``c,'' with templates of DB and M in red and green, respectively.}
\label{fig:indi}
\end{figure}

A total of 176 pure DB spectra from the literature are omitted in our catalog.
Most of them (96 spectra) have no apparent He {\small \bf I} lines or only one possible He {\small \bf I} line so that our method could not be used to recognize them.
We present one example in Figure \ref{fig:indi}; SDSS Plate--MJD--Fiber (p--m--f), 0804--52286--0262.
Another 38 spectra are generally of very poor quality, and the hyper-planes in the feature spaces can be ineffective.
Besides, an incorrect radial velocity (RV) may lead to a failure of recognition, which includes 33 of these spectra.
As for measuring RV of A type star using DB templates, there are no He {\small \bf I} lines in A-type star spectra and only He {\small \bf I} lines in DBWD spectra, the errors of RV arise
because Balmer lines in the A-type star spectrum are being identified with He {\small \bf I} lines in the DB spectrum.

Nine spectra are supposed to be DAB instead of DB.
It is worth mentioning that some spectra, for example SDSS J092604.91+264225.0, which is mis-classified as DA in \cite{2013ApJS..204....5K}, are typical DBWDs.
More details can be found in Table \ref{final:catalog} and the online table.
Table \ref{tab:etable} lists the columns of data provided in our online catalog, Table \ref{final:catalog}.

Furthermore, we add 15 pure DB spectra of 14 objects in the SDSS DR14.
Consider the number of DB objects presented in the literature and our catalog in SDSS DR12; the completeness of our ML method should be about 96.0\%.
Strong noises at the wave range of features may cause mis-classifications in this paper, which is the main disadvantage of SVM that is in need of improvement.

In general, Table \ref{tab:dbtype} lists the numbers of each types of DBWD identified in this study.

\begin{table}
\caption{Numbers of identified DBWD types.}
\label{tab:dbtype}
\centering
\begin{threeparttable}
\begin{tabular}{crr} \hline
Type	&	No. of Objects	&	No. of Spectra	\\ \hline
DB		&	1895			&	1395			\\
DB+M\tnote{a}	&	89		&	79				\\
DB:DC	&	23				&	21				\\
DBA\tnote{b}	&	627		&	465				\\
DBO		&	23				&	18				\\
DBQ		&	5				&	4				\\
DBZ		&	112				&	81				\\ \hline
\end{tabular}
\begin{tablenotes}
\item[a] The subtype and RV of the M companion can be found in the online table.
\item[b] Some of the DBA are actually DBAZ or DBAQ; they are all counted in ``DBA.''
\end{tablenotes}
\end{threeparttable}
\end{table}

\subsubsection{ID of literature}
\label{sec:idlit}

In the last column of Table \ref{final:catalog}, the numbers represent the IDs of specific literatures, which are listed as follows.  

0: firstly reported in this paper;
1: \cite{2013ApJS..204....5K}; 
2: \cite{2015A&A...583A..86K}; 
3: \cite{2015MNRAS.446.4078K}; 
4: \cite{2016MNRAS.455.3413K}; 
5: \cite{2007ApJ...664...53A}; 
6: \cite{2006ApJS..167...40E}; 
7: \cite{2004MNRAS.349.1397C}; 
8: \cite{2001MNRAS.322L..29C}; 
9: \cite{2010MNRAS.402..620R}; 
10: \cite{2008AJ....135..785W}; 
11: \cite{2011MNRAS.417.1210G}; 
12: \cite{2005RMxAA..41..155S}; 
13: \cite{2000A&AS..147..169B}; 
14: \cite{2015MNRAS.448.2260G}; 
15: \cite{2015MNRAS.446..391L}; 
16: \cite{2014ApJS..213....9D}; 
17: \cite{2011MNRAS.410.2095V}; 
18: \cite{2010ApJ...708..456R}; 
19: \cite{2013MNRAS.429.2143C}; 
20: \cite{2012AJ....143....6J}; 
21: \cite{2012ApJ...749..154G}; 
22: \cite{2011ApJ...737...28B}; 
23: \cite{2010ApJ...722..725Z}; 
24: \cite{2007A&A...470.1079V}; 
25: \cite{2005A&A...432.1025K}; 
26: \cite{1999ApJS..121....1M}; 
27: \cite{2000BaltA...9..485B}; 
28: \cite{1998BaltA...7..355B}; 
29: \cite{2005AJ....129.1483L}; 
30: \cite{1999PASP..111.1099S}; 
31: \cite{2003AJ....126.1455S}; 
32: \cite{2017A&A...600A..73C}; 
33: \cite{2004ApJ...607..426K}.

\subsection{Noteworthy Individual Objects}

Panel ``b'' in Figure \ref{fig:indi} shows the AM Canum Venaticorum (AM CVn) type spectrum (p--m--f 7324--56935--0828) that we spectroscopically identified for the ML method that only requires the intensity of the change.
The AM CVn binaries are a rare ultra-compact double degenerate system and only 43 such objects are known \citep{2015MNRAS.452.1060C,2015MNRAS.446..391L}.

We provide 66 DB spectra with M type stars as companions.
However, there are more than 30 DB M double stars in the literature \citep{2013ApJS..204....5K,2015MNRAS.446.4078K,2016MNRAS.455.3413K}.
The reason why these double stars cannot be discovered by our method is that the flux of M exceeds that of DB, which could lead to much weaker features of the DB in a spectrum.
After a visual inspection, we select 23 ``DB+M'' double stars with relative good qualities from the literature.
For these 89 double stars, we provide subtype and RV of the M companion in Table \ref{final:catalog}.
One DB+M spectrum (p--m--f 1057--52522--0613) together with templates of DB and M are illustrated in the panel ``c'' of Figure \ref{fig:indi}.

\subsection{Parameter Measurement}
\cite{2015A&A...583A..86K} has provided and analyzed parameters of DBWDs with a theoretical model.
Besides the selection of DB samples and research on the ML algorithm, 
we also provide the parameters of newly discovered DB spectra based on DB parameter templates provided by \cite{2015A&A...583A..86K}.
With the method of full spectral template matching mentioned in Section \ref{sec:datatrain}, 
we measure $T_{\text{eff}}$ and  log $g$ on several He {\small \bf I} lines, and presented the results in Table \ref{final:catalog}.
The average errors of $T_{\text{eff}}$ and log $g$ are 30.1\% and 10.6\%, respectively.

\subsection{T$_{\text{eff}}$ and Ultra-violet Color}
\begin{figure}
\centering
\includegraphics[width=0.48\textwidth]{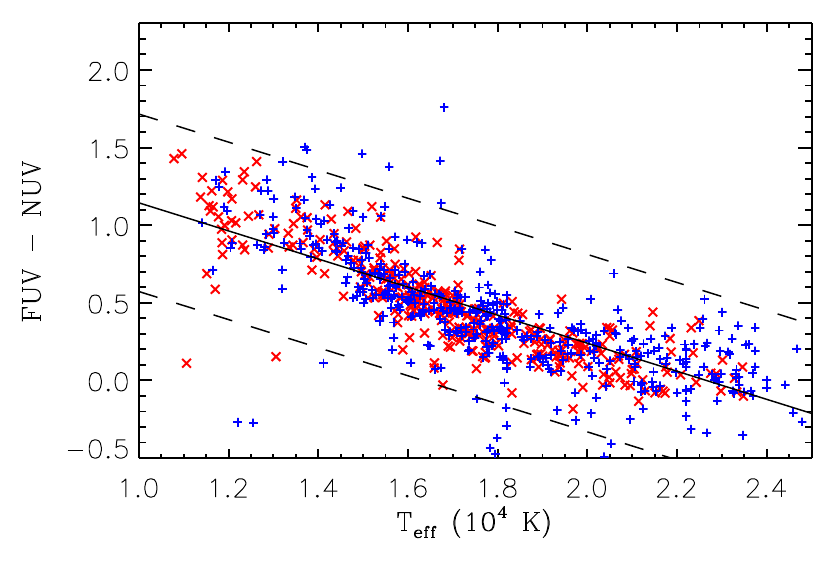}
\caption{Magnitude of $\text{FUV} - \text{NUV}$ as a function of $T_{\text{eff}}$. 
The red x-marks represent objects from the literature, while the blue plus symbols are from our catalog.}
\label{ana:uv}
\end{figure}

WDs are a type of stars that have strong intensities in the ultraviolet waveband.
The $\text{FUV} - \text{NUV}$ color from the Galaxy Evolution Explorer (GALEX) is almost reddening-free \citep{2017ApJS..230...24B}. 
All DB objects with photometric data of GALEX are cross-matched and those objects with errors in both FUV and NUV less than 0.3 mag are selected.
From Figure \ref{ana:uv}, we conclude that $T_{\text{eff}}$ and $\text{FUV} - \text{NUV}$ color are roughly linear using Equation \ref{equ:fnuv}.
The fitting variance is $\sigma\approx0.19$. 
\begin{equation}
\label{equ:fnuv}
y=-0.90\times 10^{-4}x+2.05,
\end{equation}
where $x$ is $T_{\text{eff}}$ and y is the $\text{FUV} - \text{NUV}$ color.
The majority of sources fall within the $\pm 3\sigma$ region of Equation \ref{equ:fnuv}, which is illustrated by the red dashed line in Figure \ref{ana:uv}. 

\section{Conclusion and Discussion}
\label{sec:conclusion}

We have spectroscopically identified 1999 DBWDs in the SDSS DR12 and DR14, including 58 newly identified objects, using ML, i.e., LASSO and SVM.
A total of 176 DB objects from the literature are not included in our catalog, $T_{\text{eff}}$ mostly varies around 11,000 K, and log $g$ is fixed at 8.0.
The DB spectra in this parameter range have almost no He {\small \bf I} lines; hence, our method failed to identify these spectra. 

Features of DB versus several other types of spectra were also extracted by LASSO using this procedure.
Although we cannot guarantee the completeness of our samples, we have proposed a significant scheme to extract linearly supporting features from spectra to identify DBWDs.

\begin{figure*}
\centering
\includegraphics[width=0.98\textwidth]{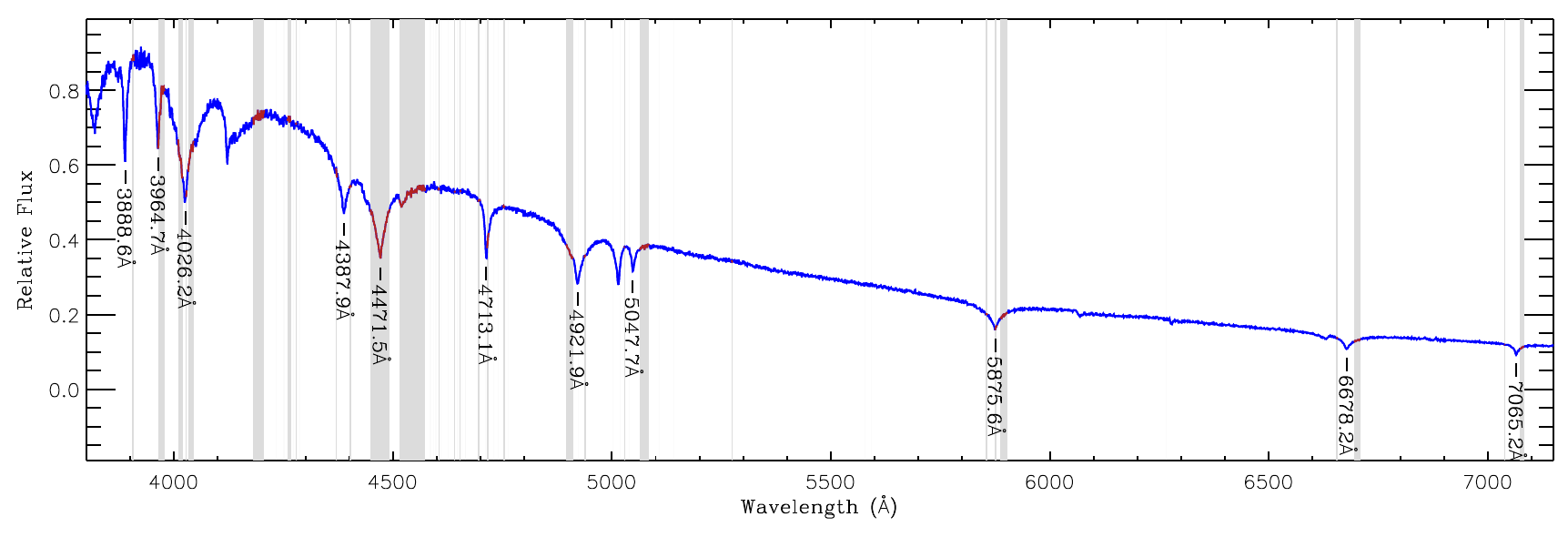}
\caption{Features of DB.
The flux at the wavelength of unique features of DB are shown in red, while the others are shown in light gray.
The He {\small \bf I} lines that contain any features are labeled.}
\label{fig:dbfeature}
\end{figure*}

Furthermore, we define all of the features illustrated in Figure \ref{fig:dbfeature}, and Tables \ref{tab:finalfeature1} and \ref{tab:finalfeature2}. 
A DB spectrum with high S/N is plotted in blue, and the features in red with a light gray background.
The features within the area of most of the He {\small \bf I} lines are asymmetric about the two sides, illustrating the characteristic of He {\small \bf I} lines in a DB spectrum.
These extracted features are demonstrated in Table \ref{tab:finalfeature1}, and do not differ from one or two atomic lines; 
the spectral line's name and position are also given in this table.
We consider the flux in the range 4693.5 -- 4699.0\AA\ as a characteristic of He {\small \bf I} 4713.1\AA\, although it appears to go beyond the range of this spectral line
because the flux of a DB spectrum begins to decrease slightly in this part.
This kind of tiny variation is usually overlooked by the human eye, but is observable when using programs to carry out the classifications.
More positions of other features are also defined in a similar fashion.
The width of the feature to the left of some spectral lines, such as He {\small \bf I} 5875.6\AA, is smaller than that on the right.
On the other side, many features that are not linked to any specific spectral lines are present in Table \ref{tab:finalfeature2}.
These features are purely data based or due to the residual sky background.

Finally, we measure the parameters, $T_{\text{eff}}$ and log $g$, of DBWDs using DB templates from \cite{2015A&A...583A..86K}. 
The consistency of $T_{\text{eff}}$ of DB objects between \cite{2015A&A...583A..86K} and our catalog is demonstrated through the $\text{FUV} - \text{NUV}$ colors from GALEXY.
The distribution of mag\_g also indicates capability of our method of searching for DBWDs in fainter objects.

\begin{table}
\caption{Features of DB located near familiar spectral lines.}
\label{tab:finalfeature1}
\centering
\begin{threeparttable}
\begin{tabular}{crrc} \hline
\multicolumn{1}{c}{He {\small \bf I} Line (\AA)}	&	\multicolumn{1}{c}{Wavelength (\AA)}	&	\multicolumn{1}{c}{Feature ID\tnote{a}}	&	Level\tnote{b}	\\	\hline
3888.6					&	3903.0 -- 3908.4	&	r144He I	&	5						\\ \hline
3964.7					&	3963.6 -- 3978.3	&	cHe I	&	4						\\ \hline
\multirow{3}{*}{4026.2}	&	4009.5 -- 4019.8	&	l64He I	&	\multirow{3}{*}{4, 5}	\\
						&	4026.2 -- 4029.0	&	cHe I	&							\\
						&	4031.8 -- 4044.8	&	r56He I	&							\\ \hline
4120.8					&	4180.2 -- 4206.3	&	r594He I	&	4, 5					\\ \hline
\multirow{2}{*}{4387.9}	&	4369.1 -- 4372.2	&	l157He I	&	\multirow{2}{*}{4}		\\
						&	4399.4 -- 4404.6	&	r115He I	&							\\ \hline
\multirow{2}{*}{4471.5}	&	4447.3 -- 4492.6	&	cHe I	&	\multirow{2}{*}{4, 5}	\\
						&	4513.3 -- 4574.1	&	r418He I	&							\\ \hline
\multirow{6}{*}{4713.1}	&	4693.5 -- 4699.0	&	l141He I	&	\multirow{6}{*}{4, 5}	\\
						&	4711.9				&	l12He I	&							\\
						&	4714.1 -- 4718.5	&	r10He I	&							\\
						&	4729.3			 	&	r162He I	&							\\
						&	4739.1 -- 4740.2 	&	r260He I	&							\\
						&	4750.1 -- 4755.5	&	r370He I	&							\\ \hline
\multirow{3}{*}{4921.9}	&	4894.4 -- 4911.4	&	l105He I	&	\multirow{3}{*}{4}		\\
						&	4922.6 -- 4923.8	&	cHe I	&							\\
						&	4935.1 -- 4940.8	&	r132He I	&							\\ \hline
\multirow{3}{*}{5015.7}	&	5001.5				&	l142He I	&	\multirow{3}{*}{4, 5}	\\
						&	5017.6 -- 5018.8	&	r19He I	&							\\
						&	5026.9 -- 5029.2	&	r112He I	&							\\ \hline
\multirow{2}{*}{5047.7}	&	5038.4 -- 5039.7	&	l80He I	&	\multirow{2}{*}{5}		\\
						&	5062.9 -- 5085.1	&	r152He I	&							\\ \hline
\multirow{3}{*}{5875.6}	&	5851.9 -- 5857.4	&	l182He I	&	\multirow{3}{*}{4}		\\
						&	5873.5 -- 5879.0	&	cHe I	&							\\
						&	5885.7 -- 5903.4	&	r101He I	&							\\ \hline
\multirow{2}{*}{6678.2}	&	6652.7 -- 6657.4	&	l208He I	&	\multirow{2}{*}{4}		\\
						&	6692.6 -- 6709.7	&	r144He I	&							\\ \hline
\multirow{2}{*}{7065.2}	&	7035.5 -- 7038.9	&	l263He I	&	\multirow{2}{*}{4}		\\
						&	7071.3 -- 7082.8	&	r69He I	&							\\ \hline
\multirow{2}{*}{7281.4}	&	7279.5				&	l19He I	&	\multirow{2}{*}{4}		\\
						&	7289.5				&	r81He I	&							\\ \hline
\end{tabular}
\begin{tablenotes}
\item[a] Feature ID is defined to demonstrate the relations between features and spectral lines. For example, cHe {\small \bf I} represents the line center, and  l123He {\small \bf I} and r123He {\small \bf I} represent the left and right 12.3\AA\ to the line center, respectively.
\item[b] Detailed scale of DB spectral wavelet decomposition, in which features can be detected.
\end{tablenotes}
\end{threeparttable}
\end{table}

\begin{table}
\caption{Features of DB not located near the familiar spectral lines.}
\label{tab:finalfeature2}
\centering
\begin{tabular}{cc} \hline
Wavelength (\AA)												&	Length (\AA)	\\	\hline
4240.3, 4252.1, 4273.7, 4281.5, 4288.5	&	---		\\
4585.6, 4587.8, 4593.0, 4596.2, 4599.4	&	---		\\
4608.9, 4647.3, 4648.4, 4658.0, 4665.5	&	---		\\
4666.6, 4763.2, 5107.4, 5141.6, 5578.3	&	---		\\
5592.4, 6264.7, 7184.6, 7340.1, 8830.8	&	---		\\
4232.5 -- 4233.5						&	1.0	\\
4257.9 -- 4266.8						&	8.9	\\
4277.6 -- 4279.6						&	2.3	\\
4602.5 -- 4605.8						&	3.3	\\
4624.9 -- 4625.9						&	1.0	\\
4638.7 -- 4642.0						&	3.3	\\
4651.5 -- 4654.8						&	3.3	\\
5272.3 -- 5277.0						&	4.7	\\ \hline
\end{tabular}
\end{table}

\acknowledgments

The authors would like to thank Drs. Hai-Feng Yang, Peng Wei and Zhen-Ping Yi for valuable discussion, and also appreciate Dr. Anthony E Lynas-Gray for the helpful suggestions. 
This work was funded by the National Basic Research Program of China (973 program, 2014CB845700) and National Natural Science Foundation of China (Grand No. 11390371/4).
This work is supported by the Astronomical Big Data Joint Research Center, co-founded by the National Astronomical Observatories, Chinese Academy of Sciences and the Alibaba Cloud.
The SDSS-III web site is \url{http://www.sdss3.org/}. SDSS-III is managed by the Astrophysical Research Consortium for the Participating Institutions of the SDSS-III Collaboration including the University of Arizona, the Brazilian Participation Group, Brookhaven National Laboratory, Carnegie Mellon University, University of Florida, the French Participation Group, the German Participation Group, Harvard University, the Instituto de Astrofisica de Canarias, the Michigan State/Notre Dame/JINA Participation Group, Johns Hopkins University, Lawrence Berkeley National Laboratory, Max Planck Institute for Astrophysics,
Max Planck Institute for Extraterrestrial Physics, New Mexico State University, New York University, Ohio State University, Pennsylvania State University, University of Portsmouth, Princeton University, the Spanish Participation Group, University of Tokyo, University of Utah, Vanderbilt University, University of Virginia, University of Washington, and Yale University.

\bibliography{ref1}



\end{document}